\documentclass[useAMS,usenatbib]{mn2e}
\usepackage{graphicx}

\def\la{\raise.5ex\hbox{$<$}\kern-.8em\lower 1mm\hbox{$\sim$}}
\def\ma{\raise.5ex\hbox{$>$}\kern-.8em\lower 1mm\hbox{$\sim$}}

\def\msol{M$_{\odot}$ }
\def\Lsol{L$_{\odot}$ }
\def\kms{$\rm km\, s^{-1}$}
\def\cm3{$\rm cm^{-3}$}
\def\Ts{$\rm T_{*}$~}
\def\Vs{$\rm V_{s}$~}
\def\n0{$\rm n_{0}$}
\def\B0{$\rm B_{0}$}

\def\Te{$\rm T_{e}$}

\def\erg{$\rm erg\, cm^{-2}\, s^{-1}$}

\def\LIR{L$_{IR}$~}
\def\L12{L$_{12\mu m}$~}
\def\F12{F$_{12\mu m}$~}

\def\Hb{H${\beta}$~}
\def\Ha{H${\alpha}$~}
\def\Hg{H${\gamma}$~}

\def\RO3{R$_{[OIII]}$}

\title[AGN and SB coexistence in galaxies]{Exploring  AGN - starburst coexistence in  galaxies at \\
z$\sim$ 0.8 by the [OIII]4959+5007/[OIII]4363 line ratio
}
\author[M. Contini]{ M. Contini 
\\
School of Physics and Astronomy, Tel Aviv University, Tel Aviv
69978, Israel \\
}

\begin{document}


\pagerange{\pageref{firstpage}--\pageref{lastpage}} \pubyear{2009}

\maketitle

\label{firstpage}

\begin{abstract}
We analyse by detailed modelling  the  spectra  observed from  the sample  galaxies  at z$\sim$0.8
presented by Ly et al (2015), constraining the models by 
 the [OIII]5007+4959/[OIII]4363 line ratios. Composite models (shock + photoionization) are adopted.
Shock velocities $\geq$ 100 \kms and preshock densities \n0$\sim$ 200 \cm3  characterize the gas surrounding
the starburst (SB),  while  \n0 are higher
by a factor of 1.5-10   in the AGN emitting gas.  SB effective  temperatures are similar to those
of quiescent galaxies (\Ts$\sim$4-7 10$^4$ K). Cloud geometrical thickness in  the SB
are $\leq$ 10$^{16}$ cm indicating  major fragmentation, while  in  
 AGN  they reach $>$ 10 pc.  O/H  are about solar for all the objects,
except for a few AGN clouds with O/H=  0.3 -0.5 solar.
Starburst models reproduce most of the data within the observational  errors.
About half of the object spectra are well fitted  by  an  accreting AGN.
Some galaxies show multiple radiation sources, such as  SB+AGN, or a double AGN. 
\end{abstract}


\section{Introduction}

Starburst (SB) - active galactic nucleus (AGN) connection
  is generally explored  in order to understand galaxy formation, structure and evolution.
Processes which  stimulate starbursts may lead also to AGN onset (Hopkins et al  2011).
 Dixon \& Joseph (2011)
 using a variety of spectroscopic diagnostics  found  that $\sim$50\% of the  luminous infrared galaxies
(LIRG) in their sample shows some evidence for an AGN. 
The luminosity of  $\sim$ 17 \% of  the sample galaxies  seems dominated by emission from AGN, 
and the remaining $\sim$ 80 \% have luminosities dominated by SB.  $\sim$50\% of the sample  galaxies
indicates  coupled AGN and starburst activities  suggesting  that  AGNs and starbursts commonly coexist. 
Contini \& Contini (2007) investigating a sample of LIRG (\LIR $>$10$^{11}$\Lsol), ultra-LIRG (ULIRG,
\LIR$>$10$^{12}$\Lsol) and hyper-LIRG (HLIRG,\LIR$>$ 10$^{13}$\Lsol) 
concluded that half of the LIRG 
contains an AGN, at least  one AGN is found in all the ULIRG and none in the HLIRG.
The connection between the AGN and the starburst is not direct and it is  affected by a (viscous) time lag  
of gas flowing 
through the AGN accretion disc leading to AGN activity delay in star formation activity
(Blank \& Duschl 2013).
AGN feedback may terminate star formation in the host galaxy  poor gas phase  and trigger it
in the rich phase (Zubovas et al. 2013).
 SB-AGN coexistence in local galaxies has 
 been investigated   by the analysis of   mid-infrared  lines
 and of the  continuum  spectral energy distribution (SED)
adopting diagnostic diagrams (e.g. Dixon \& Joseph 2011) and by the optical- UV  lines and continuum SED
in relatively high z galaxies  by detail modelling (Contini 2015 and references therein). 

In this paper we  revisit the spectra  from galaxies at z$\sim$0.8, 
presented by the Ly et al (2015) DEEP2 survey.
The observations were done by the DEIMOS multi-object spectrograph on the Keck II telescope.
The observed spectra   account for 
the [OIII]5007+ (the + indicates that the 5007 and 4959 lines are  summed), [OIII]4363, [NeIII] 3869+, 
[OII]3727+, and \Hb  lines
which  can be used to constrain the models. 
The observed spectra are generally reproduced by detailed modelling. 
We have noticed by modelling   the spectra observed from many different galaxy types (Contini 2016 and references therein),
 that  in some cases, even when the [OIII]5007+/\Hb and [OII]3727+/\Hb line ratios
are well fitted by the models,  the [OIII]4363/\Hb  systematically disagree.
For instance, [OIII] 4363/\Hb line ratios calculated by shock dominated models overpredict
the data in super-luminous SN host galaxies (Contini 2016). Therefore, the [OIII]4363/\Hb line ratios,
when available, are useful to constrain the models.
Selecting the   AGN,  SB, and/or only shock dominated models best fitting the data,
the galaxy types can be distinguished.

The  \RO3 ([OIII]5007+4959/[OIII]4363) line ratios are  favourite in the modelling process  
because they  depend only indirectly on the O/H relative abundance.
Actually,  O/H variations affect the cooling process in the recombination zone of the emitting clouds
leading to  different line spectra.
We use models  which account for both the photoionizing flux from a
 primary radiation source  and   shocks.
Shocks   yield  fragmentation of matter  by turbulence created near the shock-front and
 compression of the gas downstream,   leading to high densities which  can  trigger star formation.
Composite  models resolved the problem regarding \RO3 in AGN and LINER (low-ionization nuclear emission-line region)
spectra,  which indicated relatively high temperatures ($>$ 10$^5$K) and  densities ($>$ 10$^5$ \cm3)
in the emitting gas (Contini \& Aldrovandi 1986). 
Moreover, comparing calculated with observed line ratios the gas motion direction can be determined, i.e.
 infalling towards the radiation source or  ejected outwards (Fig. 1).
This is an important issue in view of the  SB - AGN  feedback
in galaxies.
 The calculation method is described in Sect. 2.
 Model results are  compared with the data in Sect. 3.
Discussion and concluding remarks appear in Sect. 4.

\section{Calculation of the spectra}

\begin{figure}
\centering
\includegraphics[width=8.0cm]{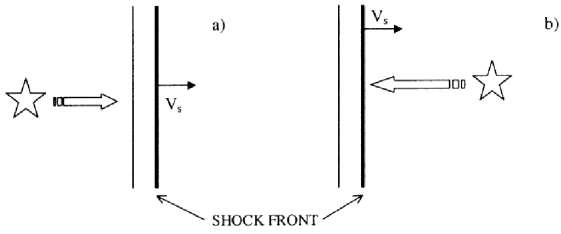}
\caption{Case a) : the cloud  withdraws from the  radiation source which is represented by the star;
 case b) the cloud approaches the radiation source}
\end{figure}

\begin{figure}
\centering
\includegraphics[width=8.0cm]{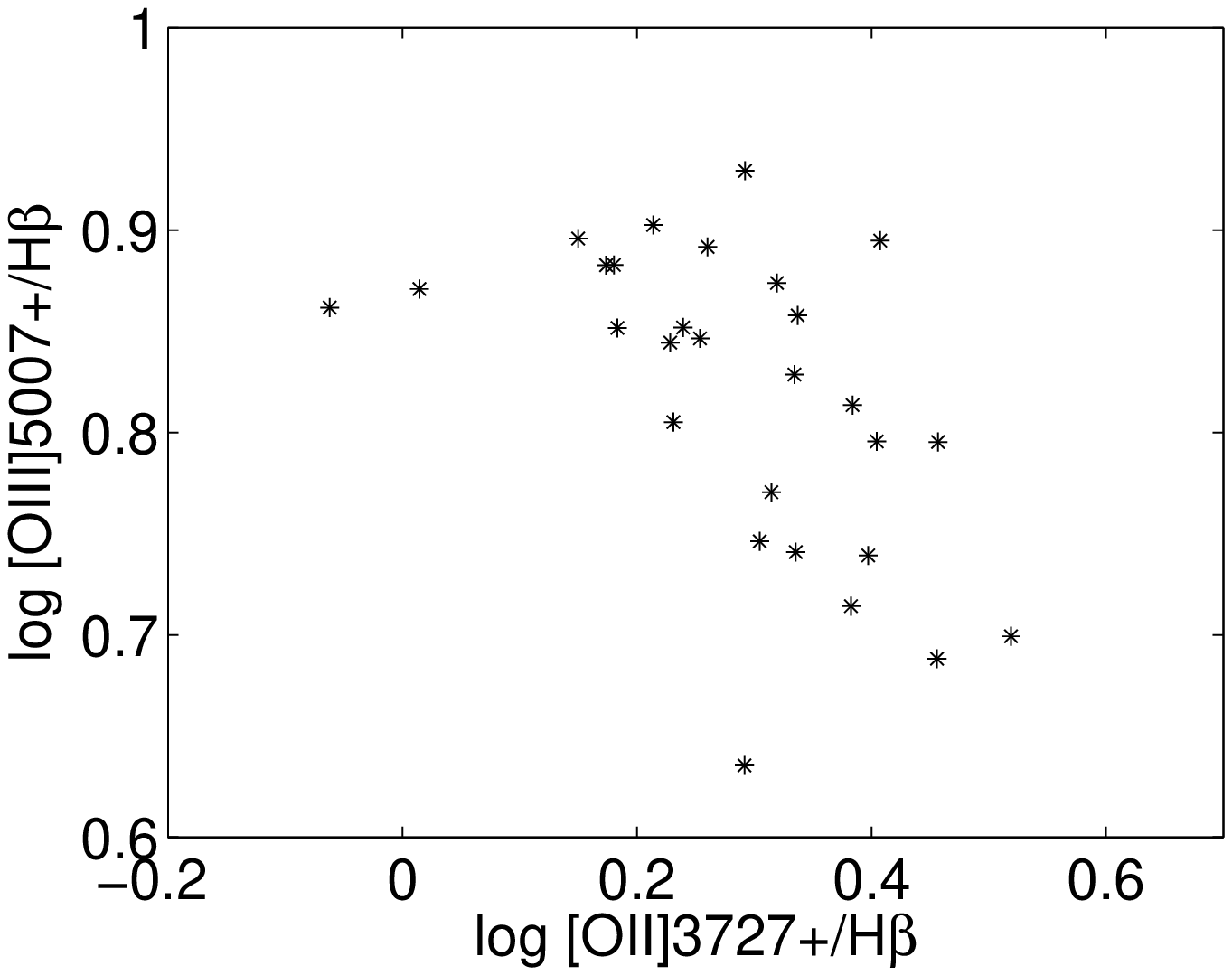}
\includegraphics[width=8.0cm]{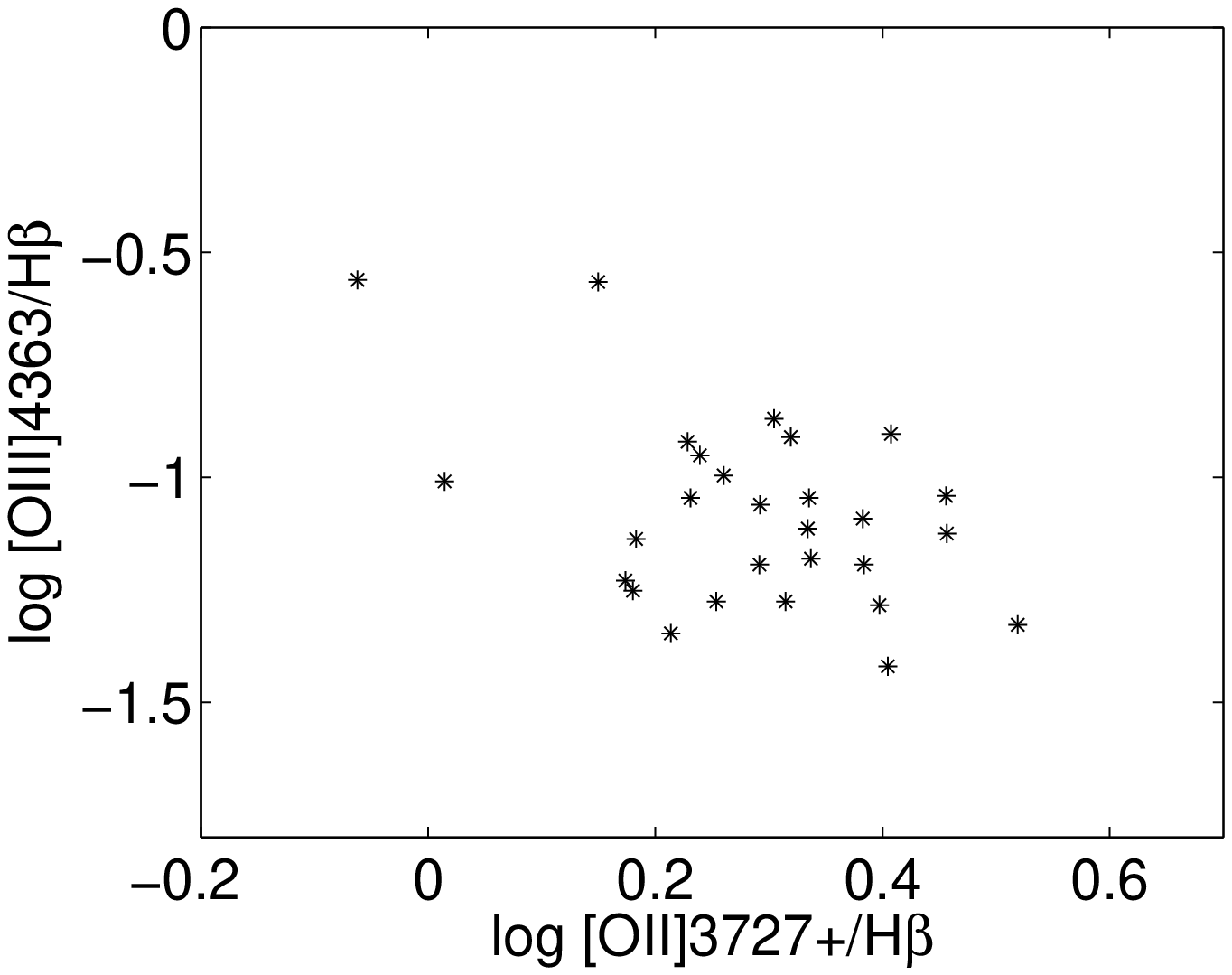}
\caption{Top : log([OIII]5007+/\Hb) (obs) vs log([OII]3727+/\Hb) (obs);
bottom : log([OIII]4363)/\Hb (obs) vs log([OII]3727+/\Hb) (obs)
}
\end{figure}

Spectroscopic data   provide   a full physical and chemical picture for local galaxies. At high redshifts
the  data are reduced to a few significant lines,  but the surveys contain hundreds of objects.
Therefore,  in order to obtain the O/H metallicity,
the oxygen line ratios are generally calculated
by the "direct methods" (see e.g. Ly et al 2015, Modjaz et al 2008).
The  "direct" or \Te ~ method (Seaton  1975, Pagel et al. 1992, etc)  was used to obtain  O/H from
the observed oxygen  to \Hb line ratios.
By this method, the ranges of the  gas physical  conditions are   chosen among those   most  suitable to the
observed line ratios.
The [OIII]4363 line  derives from the upper $^1S$ level, while [OIII] 4959 and 5007  lines derive from the 
lower $^1D$ level.
The relative rates of excitation of the $^1S$ and $^1D$ levels depend very strongly on the temperature T (Osterbrock 1974). 
Therefore, T  is calculated from the   [OIII]5007+/[OIII] 4363 ratio,
which depends also on the gas density. This  is  generally obtained from the [SII] 6717/6730 line ratios.
 In the present spectra the [SII]6717,6731
lines were not observed.
Moreover, by "direct methods"  a unique temperature is adopted throughout the whole galaxy,
neglecting the physical conditions  less adapted to the strongest lines.
 This  leads to  discrepancies between metallicities (in terms of the O/H relative abundance)
calculated by  "direct methods" and detailed modelling (see, e.g. Contini 2014a), because
the gas cools and recombines in the regions far from the photoionization source, as well as 
downstream  of  shock-fronts. By detailed modelling the gas  at low (T$<$ 10$^4$K) 
 temperatures contributes to  the line emission.  
The  line fluxes which result from integration
throughout  regions of gas at  various temperatures  are different   from those 
 calculated by an homogeneous temperature. 
If the  fractional abundances of the corresponding ions are low,
 relatively high O/H are needed to reproduce strong oxygen observed lines.
So, the metallicities calculated by  detailed modelling are generally higher than those calculated
by "direct methods".
A comparison between the results obtained by  "direct methods" and   detailed modelling  
(Contini 2014a)  demonstrates that "direct methods"  lead to O/H lower limits.

Detailed modelling  of  the spectra by pure photoionization models 
(e.g. by the CLOUDY code)  gives   satisfying results for intermediate ionization level lines.
However, galaxies at high redshifts  often  originate from mergers and show a disturbed hydrodynamic structure.
Collisional phenomena are critical in the calculation of the spectra.
We suggest that
models based on the coupled effect of photoionization and shocks
 are the closest approximation to the  complex  structure of the emitting gas.
Therefore, the  SUMA code (Contini 2015 and references therein) is used. 

The code simulates the physical conditions in an emitting gaseous cloud under the coupled effect of
photoionization from   the primary radiation source ( SB or AGN) and shocks. The line and continuum emission
from the gas are calculated consistently with dust-reprocessed radiation in a plane-parallel geometry.
 The calculations start at the shock front where the gas is compressed and thermalized adiabatically,
reaching the maximum temperature in the immediate post-shock region
($T(K)\sim 1.5\times 10^5/(V_{\rm s}/100$ km s$^{-1}$$)^{2}$,
where \Vs is the shock velocity).
T decreases  downstream following the cooling rate.
The input parameters such as  \Vs, the atomic preshock density \n0 and
the preshock magnetic field \B0 (for  all models \B0=10$^{-4}$Gauss is adopted)
define the hydrodynamical field. 

 The input parameters  that represent the primary radiation 
for a   SB are  the effective temperature \Ts
 and the ionization parameter $U$.
For an AGN, the primary radiation is the power-law radiation
flux  from the active center $F$  in number of photons cm$^{-2}$ s$^{-1}$ eV$^{-1}$ at the Lyman limit  and
spectral indices  $\alpha_{UV}$=-1.5 and $\alpha_X$=-0.7.
  The primary radiation source 
is independent  but it affects the surrounding gas.
In contrast, the secondary diffuse radiation is emitted from the slabs of gas heated
by the radiation flux reaching the gas and, in particular, by the shock.
In our model the gas region surrounding the radiation source is not considered as a
unique cloud, but as an ensemble of fragmented filaments. The geometrical
thickness of these filaments is  an input parameter of the code (D) which is
calculated consistently with the physical conditions and element abundances of
the emitting gas.
Primary  and  secondary radiations are  calculated by radiation
transfer throughout the slabs downstream.
The fractional abundances of the ions are calculated resolving the ionization equations.
The dust-to-gas ratio ($d/g$) and the  abundances of He, C, N, O, Ne, Mg, Si, S, A, Fe, relative to H,
are also accounted for. 

In the modelling process, we   aim to reproduce the  observed line ratios  for each element.
Each line has a different  strength which  translates into  the different precision  by the fitting process.
A minimum number of significant lines ([OIII] 5007+,  [OII]3727+, [OIII]4363, [NII], \Hb, \Ha) is
necessary to constrain the model
but the  number of the observed lines  does not interfere with the modelling process.
We deal with  line ratios to avoid  distance and morphological effects.
We start  adopting solar abundances by Allen (1976) in the
 first   modelling trials because their values are between the two more recent ones by Anders \& Grevesse (1989) 
and Asplund et al (2009). 
A perfect fit of the observed line ratios is not realistic because
the observed data have errors, both random and systematic.
The uncertainty in the calculation
is due to  the atomic parameters (within 10 \%)  which are
often updated.
The strongest lines  are reproduced by $<$ 10\%, the weakest by $\sim$ 50 \%.
 The calculation code and our modelling method  are described  by Contini (2014a, 2016 and references therein).

Before starting the modelling process of the present survey galaxies, some characteristics can be guessed by
 comparing different line ratios in Fig. 2. The top diagram shows that 
 excluding three objects (ID 4, 15 and 16)  the slope shown by the line ratio band suggests that a radiation 
mechanism should be adopted. 
In the bottom diagram, excluding three objects (ID 4, 8 and 15), a well defined slope is not seen.
This reveals   large ranges of  temperatures and densities  in the emitting gas, indicating that other mechanisms 
such as e.g. shocks could be at work.

\begin{figure}
\centering
\includegraphics[width=8.0cm]{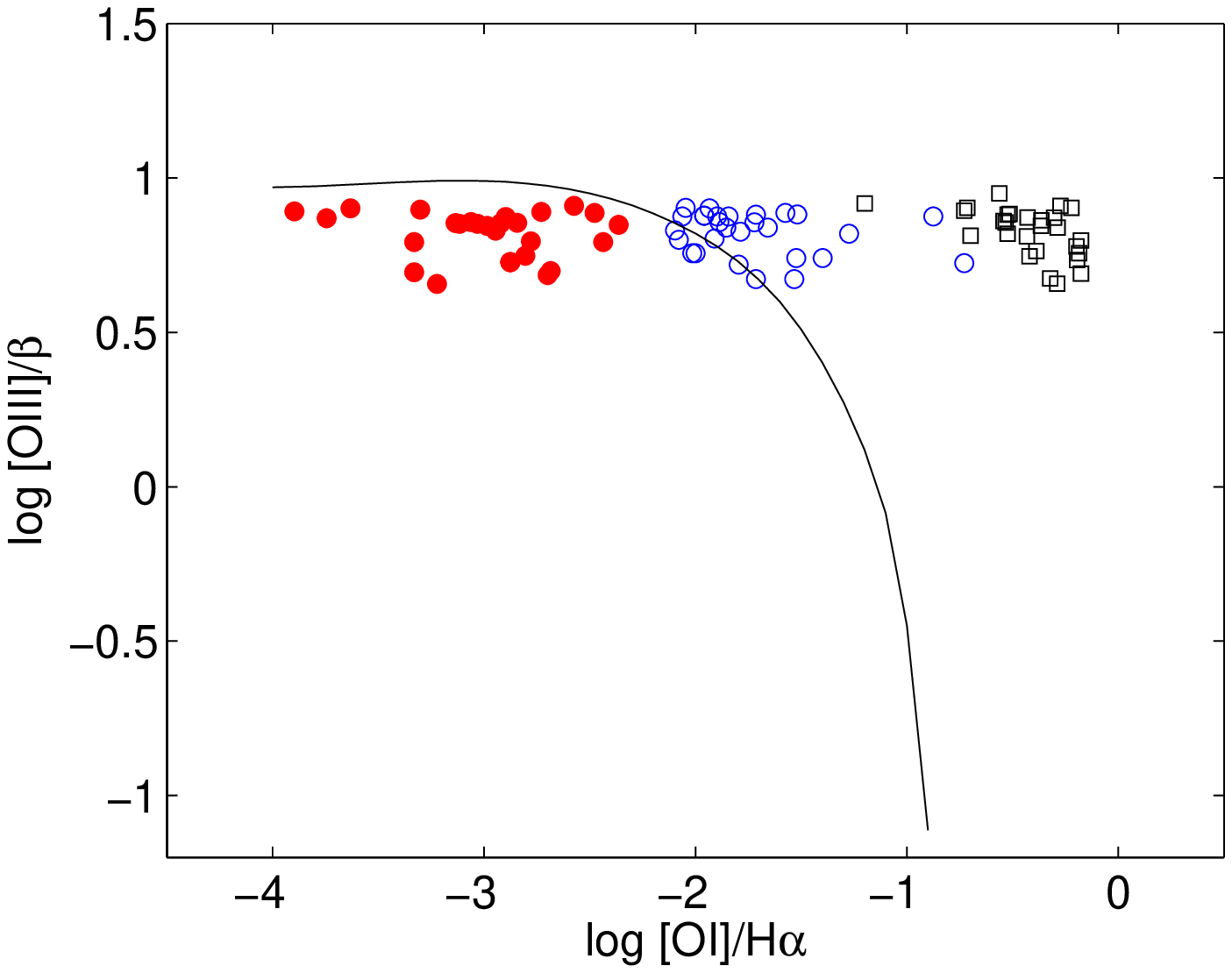}
\includegraphics[width=8.0cm]{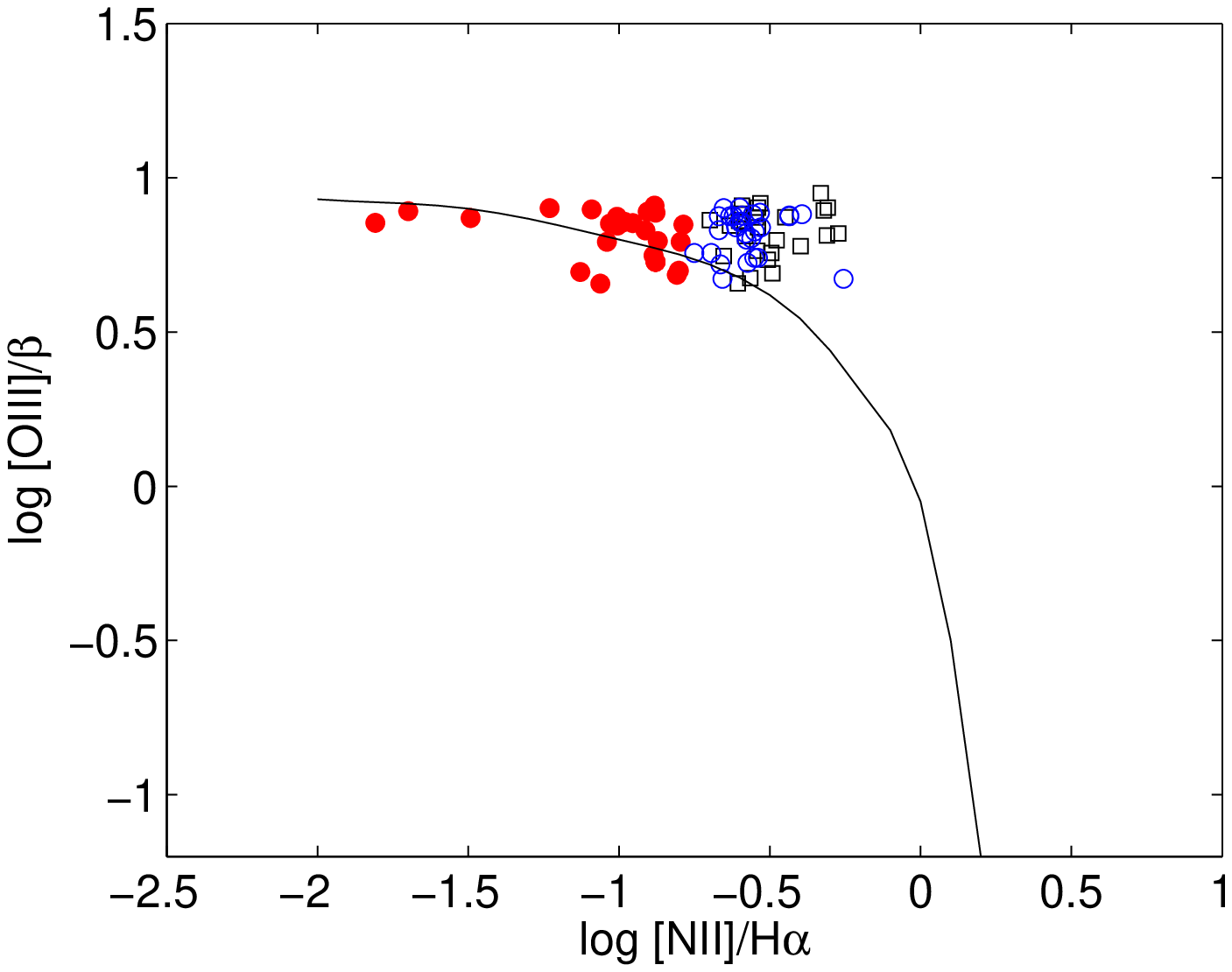}
\includegraphics[width=8.0cm]{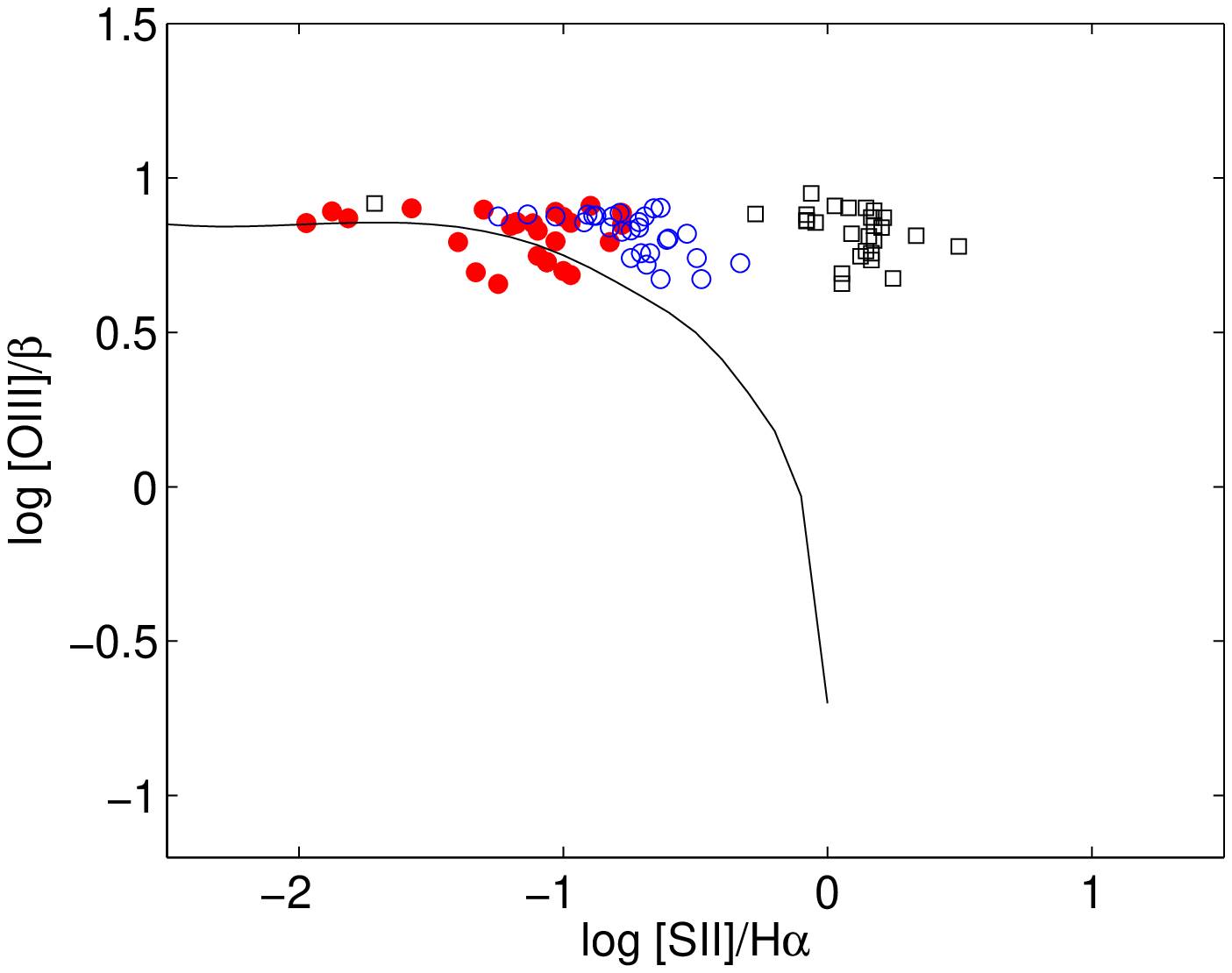}
\caption{
Comparison of  our  models with Kewley et al diagrams.
red filled circles: SB models; blue circles : AGN infalling models;  black squares :
AGN outflow models.
}
\end{figure}

\section{Modelling results}

\begin{table*}
\centering
\caption{Comparison of observed line ratios  to \Hb with model results}
\tiny{
\begin{tabular}{cccccccccccccc} \hline  \hline
\       ID & z     &FWHM$^1$     &\Hb$^2$     & [OII]3726+3728/\Hb      & [NeIII]3869/\Hb         & \Hg/\Hb                   &[OIII]4363/\Hb         &[OIII]4959+5007/\Hb (corr)  & \RO3    \\ \hline
\    1     & 0.819 & 170.9   &6.610  &2.165$_{-0.051}^{+0.101}$&0.409$_{-0.040}^{+0.040}$&0.474$_{-0.021}^{+0.019}$&0.090$_{-0.018}^{+0.018}$&5.507$_{-0.067}^{+0.156}$ (5.15) &57.27$_{-8.8}^{+11.75}$\\
\   MSB1  &  -    &  -      &  -    &1.800                  &0.300                  &0.460                  &0.100                  &5.000   &50.                        \\
\   Mpl1  &  -    &  -      &  -    &2.180                  &0.590                  &0.450                  &0.030                  &5.600   &186.66                      \\
\   Mpl01 &  -    &  -      &  -   &2.050                  &0.500                  &0.460                  &0.030                  &5.110     &170.3                     \\
\   2     &  0.749 &87.300  &5.070  &1.821$_{-0.235}^{+0.705}$ & 0.465$_{-0.075}^{+0.151}$&         -       &0.101$_{-0.020}^{+0.016}$&7.792$_{-0.351}^{+0.200}$ (6.85) & 67.84$_{-6.62}^{+19.856}$  \\
\  MSB2   &       &        &       & 1.800                  &0.561                  &0.450                  &0.200                  &7.200   &36.                          \\
\  Mpl2   &       &        &       & 1.900                  &0.700                  &0.450                  &0.045                  &7.400   &164.44                     \\
\  Mpl02  &       &        &        &1.900                  &0.400                  &0.450                  &0.041                  &6.990   &170.48                     \\
\     3   &0.710  & 109.2  &   4.510 &2.413$_{-0.103}^{+0.103}$&0.475$_{-0.043}^{+0.085}$&0.484$_{-0.029}^{+0.022}$&0.081$_{-0.023}^{+0.018}$&5.178$_{-0.085}^{+0.198}$ (4.45) &54.98$_{-10.47}^{+20.95}$\\
\  MSB3   &       &        &       &2.400                  &0.320                  &0.460                  &0.075                  &4.500     &60.                   \\
\ Mpl3    &       &        &       &2.200                  &0.600                  &0.440                  &0.050                  &5.000     &100.                   \\
\ Mpl03   &       &        &       &2.200                  &0.400                  &0.500                  &0.040                  &4.770     &119.25                   \\
\    4    & 0.788 &83.2    &3.690  &0.867$_{-0.094}^{+0.094}$& 0.187$_{-0.066}^{+0.066}$&         -        &0.275$_{-0.063}^{+0.079}$ & 7.273$_{-0.578}^{+0.385}$ (5.0) &  18.19$_{-1.692}^{+5.923}$\\
\  MSB4   &       &        &      & 1.100                  & 0.300                 &  0.450                & 0.200                  & 5.000    &25.                   \\
\  Mpl4   &       &        &      &0.990                  &0.260                  &0.440                  &0.072                  &5.120       &71.1                    \\
\  Mpl04  &       &        &      &0.820                  &0.160                  & 0.43                  & 0.450                  &5.100      &11.33                    \\
\    5     & 0.771&105.7   &19.620  &1.960$_{-0.084}^{+0.084}$ &0.552$_{-0.022}^{+0.017}$& 0.468$_{-0.008}^{+0.006}$& 0.087$_{-0.009}^{+0.007}$& 8.497$_{-0.053}^{+0.080}$ (7.87) & 90.48$_{-8.133}^{+10.167}$\\
\  MSB5   &       &        &      &  1.940                 & 0.500                &  0.450                &  0.150                  &8.000      &53.3                   \\
\  Mpl5   &       &        &      & 1.760                  &0.560                 & 0.460                 & 0.052                  &7.890       &151.7                  \\
\  Mpl05  &       &        &      & 2.000                  &0.510                 & 0.460                 & 0.310                  &8.090       &26.1                  \\
\   6     & 0.717 &99.8   &11.240 & 2.539$_{-0.095}^{+0.221}$&0.403$_{-0.020}^{+0.035}$&0.468$_{-0.007}^{+0.016}$&0.038$_{-0.005}^{+0.011}$& 6.245$_{-0.081}^{+0.081}$ (4.99) & 131.24$_{-15.91}^{+47.72}$\\
\   MSB6  &       &       &       & 2.500                  &0.350                  &0.460                 & 0.055                  &4.930      &89.6                    \\
\   Mpl6  &       &       &       & 2.460                  &0.460                  &0.443                 & 0.028                  &5.19        &185.7                   \\
\  Mpl06  &       &       &       & 2.473                  & 0.460                 &0.460                 & 0.050                  &5.200       &104.                     \\
\   7     & 0.762&108.5   &12.810 & 3.304$_{-0.260}^{+0.208}$&0.410$_{-0.025}^{+0.032}$&        -         &0.047$_{-0.012}^{+0.021}$&  5.003$_{-0.081}^{+0.040}$ (3.9) &83.49$_{-12.85}^{+25.69}$    \\
\   MSB7   &       &      &        & 3.400                  &0.420                  &0.460                &  0.170                  &4.100      &24.1                       \\
\   Mpl7   &       &       &       &3.100                  &0.420                  &0.440                 & 0.040                  &4.200       &105.                     \\       
\   Mpl07  &        &      &       &3.300                  &0.460                  &0.460                 & 0.180                  &4.100        &22.78                     \\
\   8      &0.765 &83.7   & 2.700  &1.412$_{-0.097}^{+0.779}$& 0.112$_{-0.029}^{+0.058}$&         -       & 0.272$_{-0.062}^{+0.078}$& 7.866$_{-0.168}^{+0.589}$ (6.59) &24.25$_{-4.217}^{+8.434}$\\
\   MSB8   &      &       &        & 1.860                  &0.200                  &0.460                &  0.290                  &6.500     &22.4                   \\
\   Mpl8   &      &        &       &1.300                  &0.227                  &0.444                 & 0.070                  &7.000      &100.                   \\
\   Mpl08  &      &        &       &1.540                  &0.180                  &0.460                 & 0.263                  &6.760       &25.7                  \\
\   9      &0.791&143.500  &6.360  &2.496$_{-0.079}^{+0.197}$& 0.380$_{-0.048}^{+0.040}$& 0.468$_{-0.020}^{+0.017}$& 0.052$_{-0.009}^{+0.022}$&5.485$_{-0.099}^{+0.124}$ (4.09) & 78.65$_{-14.98}^{29.964}$ \\
\   MSB9   &     &         &       & 2.400                  &0.300                  &0.460                 & 0.066                  &4.160     &59.4                      \\
\   Mpl9    &      &       &       &2.600                  &0.380                  &0.440                  &0.040                  &4.000      &100.                       \\
\   Mpl09   &      &       &       &2.100                  &0.420                  &0.460                  &0.053                  &4.500       &84.9                       \\
\  10      &0.794&109.1   &12.910  &2.159$_{-0.140}^{+0.093}$&0.435$_{-0.026}^{+0.017}$& 0.468$_{-0.01}^{+0.008}$&0.077$_{-0.014}^{+0.009}$& 6.739$_{-0.04}^{+0.094}$ (6.55) &85.11$_{-12.38}^{+15.474}$ \\
\  MSB10    &     &         &      &2.300                  &0.477                  &0.460                  &0.080                  &6.750      &84.4                   \\
\  Mpl10    &     &         &      &2.000                  &0.490                  &0.440                  &0.040                  &6.690      &167.25                   \\
\  Mpl010    &    &         &      &2.200                  &0.400                  &0.460                  &0.075                  &6.400       &85.3                   \\
\  11      &0.798 &86.3    &5.400  &2.086$_{-0.065}^{+0.065}$&0.509$_{-0.062}^{+0.035}$&0.474$_{-0.020}^{+0.014}$& 0.123$_{-0.011}^{+0.025}$ & 7.48$_{-0.118}^{0.148}$ (7.17)& 58.30$_{-11.18}^{+3.195}$  \\
\  MSB11     &     &        &       &2.100                  &0.490                  &0.460                  &0.090                  &7.200      &80.                           \\
\  Mpl11      &     &       &      &2.160                  &0.490                  &0.460                  &0.060                  &7.140       &119.                     \\
\  Mpl011     &     &       &      &2.060                  &0.530                  &0.460                  &0.086                  &7.300       &84.9                       \\
\  12      &0.749&105.5    &7.910  &1.702$_{-0.0}^{+0.229}$ &  0.482$_{-0.039}^{+0.046}$&0.468$_{-0.017}^{+0.020}$ &0.090$_{-0.009}^{+0.007}$ & 6.384$_{-0.056}^{+0.130}$ (6.39)& 70.87$_{-5.956}^{+4.756}$ \\
\  MSB12      &      &      &       &1.900                  &0.420                  &0.460                  &0.060                  &6.340       &105.7               \\
\  Mpl12      &      &      &      &1.850                  &0.470                  &0.460                  &0.050                  &6.500        &130.                \\
\  Mpl012     &      &      &      &1.700                  &0.470                  &0.460                  &0.080                  &6.760        &84.5                \\
\  13      &0.794 &91.200  &4.050  &2.862$_{-0.185}^{+0.369}$&0.459$_{-0.035}^{+0.052}$&0.468$_{-0.020}^{+0.016}$& 0.075$_{-0.015}^{+0.015}$ & 6.241$_{-0.103}^{+0.154}$ (5.58) & 74.34$_{-11.437}^{15.25}$ \\
\  MSB13      &       &     &      &2.600                  &0.460                  &0.460                  &0.090                  &6.250         &69.4                \\
\  Mpl13      &       &     &      &3.000                  &0.470                  &0.460                  &0.034                  &5.640         &165.88                 \\
\  Mpl013     &       &     &      &2.500                  &0.430                  &0.460                  &0.060                  &5.800        &96.66                 \\
\  14      &0.794&190.900 &16.990  &2.173$_{-0.089}^{+0.089}$& 0.476$_{-0.021}^{+0.017}$& 0.468$_{-0.009}^{+0.009}$&0.066$_{-0.007}^{+0.007}$& 7.209$_{-0.040}^{+0.040}$ (6.94) & 105.147$_{-9.669}^{+12.09}$\\
\  MSB14      &       &     &      & 2.300                  &0.490                 & 0.460                  &0.090                  &7.130        &79.2               \\
\  Mpl14      &       &     &      &2.120                  &0.540                  &0.455                  &0.043                  &7.020          &163.25              \\
\  Mpl014     &       &      &     &2.240                  &0.477                  &0.460                  &0.078                  &7.220          &92.56              \\
\  15      &0.755&132.8   &12.520  &1.034$_{-0.027}^{+0.096}$& 0.480$_{-0.012}^{+0.040}$&         -         & 0.098$_{-0.008}^{+0.008}$&7.428$_{-0.042}^{+0.084}$ (7.06) & 72.096$_{-6.92}^{+4.614}$ \\
\  MSB15      &       &      &      &1.000                  &0.340                  &0.460                 & 0.180                  &7.400        &41.1              \\
\  Mpl15      &       &      &     &1.000                  &0.540                  &0.440                  &0.070                  &7.200         &102.85               \\
\  Mpl015     &       &      &     &1.070                  &0.540                  &0.460                  &0.080                  &6.940         &86.75               \\ \hline

\end{tabular}}

$^1$ in \kms; $^2$ in 10$^{-17}$ \erg;

\end{table*}

\begin{table*}
\centering
\centerline{Table 1-continued}
\                       \\
\tiny{
\begin{tabular}{cccccccccccc} \hline  \hline
\       ID & z     &FWHM     &\Hb     & [OII]3726+3728/\Hb      & [NeIII]3869/\Hb         & \Hg/\Hb                   &[OIII]4363/\Hb         &[OIII]4959+5007/\Hb (corr)  & \RO3    \\ \hline
\  16       &  0.774& 205.9 &8.700  &1.958$_{-0.00}^{+0.945}$& 0.272$_{-0.059}^{+0.073}$&          -       & 0.064$_{-0.005}^{+0.041}$ &4.320$_{-0.124}^{+0.124}$ (3.4) &46.844$_{-12.49}^{+18.738}$    \\
\  MSB16      &       &      &     &2.100                  &0.230                  &0.460                  &0.090                  &3.270         &36.3             \\
\  Mpl16      &       &      &     &1.850                  &0.303                  &0.440                  &0.027                  &3.500         &129.6              \\
\  Mpl016     &       &      &     &1.600                  &0.230                  &0.460                  &0.040                  &3.870   &96.75                     \\
\  17      &0.764&132.7    &7.130  &2.420$_{-0.074}^{+0.596}$& 0.618$_{-0.048}^{+0.096}$&           -       & 0.064$_{-0.013}^{+0.016}$& 6.508$_{-0.074}^{+0.147}$ (5.73) & 89.58$_{-17.34}^{+28.898}$ \\
\  MSB17      &       &      &     & 2.500                  &0.560                  &0.460                 & 0.065                  &6.100        &93.8               \\
\  Mpl17      &       &      &     &2.420                  &0.670                  &0.440                  &0.060                  &5.930         &98.83                \\
\  Mpl017     &       &      &     &2.300                  &0.640                  &0.460                  &0.050                  &5.500         &110.                \\
\  18      &0.789&136.4    &9.300  &1.794$_{-0.194}^{+0.129}$& 0.553$_{-0.031}^{+0.072}$& 0.468$_{-0.020}^{+0.021}$& 0.053$_{-0.012}^{+0.012}$& 7.022$_{-0.113}^{+0.113}$ (6.12) & 115.5$_{-14.9}^{+37.258}$\\
\  MSB18      &       &      &      &1.800                  &0.500                  &0.460                  &0.040                  &6.200        &155.               \\
\  Mpl18      &       &      &     &1.830                  &0.630                  &0.450                  &0.036                  &6.200         &172.2               \\
\  Mpl018     &       &      &     &2.000                  &0.520                  &0.460                  &0.051                  &5.900   &115.7                      \\
\  19      &0.856&144.0  & 14.510 &1.636$_{-0.026}^{+0.021}$& 0.263$_{-0.007 }^{+0.024}$& 0.596$_{-0.015}^{+0.018}$& 0.045$_{-0.006 }^{+0.013}$ & 7.989$_{-0.132}^{+0.066}$ (5.8) & 128.95$_{-23.99}^{+17.99}$\\
\  MSB19      &       &      &      &1.800                  &0.270                  &0.460                 & 0.040                  &5.870         &146.7              \\
\  Mpl19      &       &      &     &1.470                  &0.230                  &0.440                  &0.057                  &5.850          &102.6               \\
\  Mpl019     &       &      &     &1.500                  &0.210                  &0.460                  &0.073                  &6.190          &84.93        \\ 
\  20      &0.784&108.0    &6.820  &2.017$_{-0.221}^{+0.276}$& 0.267$_{-0.023}^{+0.046}$& 0.468$_{-0.024}^{+0.030}$& 0.135$_{-0.039}^{+0.026}$& 5.575$_{-0.159}^{+0.132}$ (4.90) & 36.499$_{-5.84}^{+14.6}$\\
\  MSB20      &       &      &     &2.300                  &0.300                  &0.460                  &0.127                  &4.830          &38.              \\
\  Mpl20      &       &       &    &1.800                  &0.350                  &0.440                  &0.054                  &5.200          &96.3               \\
\  Mpl020     &       &       &    &1.940                  &0.490                  &0.460                  &0.054                  &4.960    &91.85              \\
\  21      &0.823&110.9    &5.580  &1.692$_{-0.115}^{+ 0.344}$&0.423$_{-0.046}^{+0.108}$& 0.468$_{-0.034}^{+0.032}$& 0.120$_{-0.026}^{+0.026}$& 6.989$_{-0.223}^{+0.179}$ (6.29) &52.46$_{-8.507}^{+11.343}$\\
\  MSB21      &        &       &   &1.800                  &0.470                  &0.460                  &0.135                  &6.290           &46.6              \\
\  Mpl21      &        &       &   &1.580                  &0.390                  &0.450                  &0.040                  &6.300           &157.5             \\
\  Mpl021     &        &       &   &1.900                  &0.580                  &0.460                  &0.07                   &6.300          &90.           \\
\  22      &0.784 &85.0    &6.360  &1.735$_{-0.043}^{+0.343}$& 0.429$_{-0.043}^{+0.054}$& 0.468$_{-0.021}^{+0.028}$& 0.112$_{-0.028}^{+0.008}$& 7.110$_{-0.207}^{+0.103}$ (6.9) & 61.69$_{-9.49}^{+12.654}$\\
\  MSB22      &        &       &    &1.770                  &0.450                  &0.460                  &0.120                  &7.100         &59.17             \\
\  Mpl22      &        &       &   &1.760                  &0.390                  &0.450                  &0.040                  &7.170          &179.25             \\
\  Mpl022     &        &       &   &1.900                  &0.440                  &0.460                  &0.073                  &7.000     &95.9                   \\
\  23      &0.766 &97.6    &7.670  &1.492$_{-0.00}^{+0.720}$& 0.369$_{-0.060}^{+0.100}$&          -         &0.059$_{-0.019}^{+0.015}$&  7.631$_{-0.108}^{+0.135}$ (8.5) & 144.017$_{-57.607}^{+0.0}$\\
\  MSB23      &        &       &    &1.700                  &0.480                 & 0.460                 & 0.034                  &7.700          &226.47             \\
\  Mpl23      &        &       &   &1.600                  &0.447                  &0.450                  &0.050                  &8.300            &166.           \\
\  Mpl023     &        &       &   &1.500                  &0.460                  &0.460                  &0.080                  &8.300            &97.6          \\
\  24      &0.721 &83.1   &14.890  &1.525$_{-0.076}^{+0.076}$& 0.409$_{-0.008}^{+0.027}$& 0.468$_{-0.009}^{+0.008}$&  0.073$_{-0.008}^{+0.009}$&   7.105$_{-0.057}^{+0.057}$ (6.93) & 94.92$_{-10.43}^{+8.344}$\\
\  MSB24      &        &       &   &1.520                  &0.410                  &0.460                  &0.040                  &7.160           &179.            \\
\  Mpl24      &        &       &   &1.500                  &0.400                  &0.450                  &0.043                  &7.200           &167.4            \\
\  Mpl024     &        &       &   &1.610                  &0.480                  &0.460                  &0.075                  &6.900            &92.0       \\
\  25      &0.789&102.4  &  7.330  &2.857$_{-0.336}^{+0.269}$ &0.530$_{-0.038}^{+0.077}$&  0.468$_{-0.026}^{+0.024}$ & 0.091$_{-0.015}^{+0.012}$&   4.878$_{-0.065}^{+0.130}$ (4.42) & 48.60$_{-8.526}^{+5.116}$\\
\  MSB25      &        &       &   & 2.700                  &0.450                  &0.460                  &0.073                  &4.850          &69.3             \\
\  Mpl25      &         &      &   &2.900                  &0.600                  &0.450                  &0.022                  &4.700            &213.6         \\
\  Mpl025     &        &       &   &2.500                  &0.580                  &0.460                  &0.053                  &4.700            &88.7    \\
\  26      &0.774&109.2   &15.690  &2.064$_{-0.041}^{+0.031}$& 0.309$_{-0.020}^{+0.020}$& 0.479$_{-0.011}^{+0.014}$& 0.053$_{-0.008}^{+0.006}$& 5.895$_{-0.044}^{+0.089}$ (5.78) & 109.09$_{-14.307}^{+14.307}$\\
\  MSB26      &        &       &   &2.100                  &0.400                  &0.460                  &0.043                  &6.200           &144.18          \\
\  Mpl26      &        &       &   &2.880                  &0.390                  &0.450                 & 0.044                  &6.000           &136.4            \\
\  Mpl026     &        &       &   &2.000                  &0.370                  &0.460                 & 0.057                  &5.730           &100.5       \\
\  27      &0.729& 86.4    &7.120  &2.555$_{-0.330}^{+0.220}$& 0.479$_{-0.033}^{+0.077}$& 0.468$_{-0.021}^{+0.017}$& 0.125$_{-0.012}^{+0.024}$& 7.849$_{-0.231}^{+0.132}$ (6.87) & 54.97$_{-7.715}^{+9.644}$ \\
\   MSB27     &         &       &   &2.300                  &0.460                  &0.460                  &0.136                  &6.600           &48.5           \\
\  Mpl27      &         &       &  &2.340                  &0.440                  &0.460                  &0.050                  &7.170            &143.4           \\
\  Mpl027     &         &       &  &2.440                  &0.530                  &0.460                  &0.080                  &7.000          &87.5        \\
\  28      &0.732&114.5  & 22.760  &1.515$_{-0.028}^{+0.042}$& 0.504$_{-0.009}^{+0.014}$& 0.468$_{-0.004}^{+0.006}$& 0.056$_{-0.005}^{+0.002}$& 7.633$_{-0.019}^{+0.037}$ (7.42)& 132.44$_{-5.333}^{+10.667}$ \\
\  MSB28   &     &       &         &1.500                  &0.420                  &0.460                  &0.045                  &7.500             &166.7          \\
\  Mpl28   &      &      &        & 1.540                  &0.500                  &0.440                  &0.044                  &7.600             &172.7         \\
\  Mpl028  &      &      &         &1.500                  &0.500                  &0.460                  &0.090                  &7.540      &83.78    \\  \hline

\end{tabular}}

\end{table*}

\begin{table}
\centering
\caption{Description of models MSB1-MSB28}
\small{
\begin{tabular}{lccccccccccc} \hline  \hline
   mod & \Vs  &  \n0 & \Ts     & $U$   & $D$        & O/H      &   Ne/H   & \Hb \\
\      & 1    & 2    & 3       &  4    & 5          & 6         & 6        & 7        \\ \hline
\ MSB1 & 200  & 120  &  4.4    & 4  & 1.         & 7.       & 1.       & 0.003    \\
\ MSB2 & 220  & 120  &  5.     & 4  & 0.8        & 6.5      & 1.       & 0.002   \\
\ MSB3 & 100  & 220  &  4.8    & 1.4 &0.8         & 6.5      & 1.       & 0.005   \\
\ MSB4 & 100  & 100  &  4.3    & 4.3 & 0.9        & 5.0      & 0.6      & 0.001   \\
\ MSB5 & 110  & 210  & 4.1     & 4.9 & 0.4        & 6.5      & 1.       & 0.002   \\
\ MSB6 & 100  & 220  & 5.      & 1.4 & 1.1        & 6.5      & 1.       & 0.006   \\
\ MSB7 & 100  & 220  & 4.5     & 0.9 & 0.25       & 6.3      & 1.       & 0.002   \\
\ MSB8 & 80   & 100  & 4.1     & 4.5 & 0.7        & 6.4      & 0.4      & 0.001   \\
\ MSB9 & 100  & 220  & 4.7     & 1.4 & 0.8        & 6.5      & 1.       & 0.005   \\
\ MSB10& 100  & 220  & 5.2     & 1.5 & 0.8        & 6.5      & 1.       & 0.004   \\
\ MSB11& 100  & 220  & 5.2     & 1.7 & 0.8        & 6.5      & 1.       & 0.004   \\
\ MSB12& 110  & 220  & 5.0     & 2.2 & 1.0        & 6.2      & 1.       & 0.007   \\
\ MSB13& 100  & 220  & 5.4     & 1.2 & 0.7        & 6.8      & 1.       & 0.004   \\
\ MSB14& 100  & 220  & 5.4     & 1.4 & 0.7        & 6.8      & 1.       & 0.004   \\
\ MSB15& 100  & 180  & 5.4     & 3  & 0.4        & 7.       & 0.7      & 0.002   \\
\ MSB16& 200  & 120  & 4.0     & 3.4 & 1.         & 6.6      & 1.       & 0.003   \\
\ MSB17& 130  & 220  & 4.      & 3  & 0.8        & 6.5      & 1.       & 0.008   \\
\ MSB18& 130  & 220  & 6.4     & 2  & 1.45       & 6.2      & 1.       & 0.016   \\
\ MSB19& 130  & 220  & 6.0     & 2.1 & 1.45       & 6.3      & 0.6      & 0.016   \\
\ MSB20& 100  & 250  & 5.2     & 1.1 & 0.3        & 6.3      & 0.7      & 0.003   \\
\ MSB21& 100  & 250  & 5.7     & 1.3 & 0.3        & 6.3      & 0.9      & 0.003   \\
\ MSB22& 80   & 270  & 6.      & 1.1 & 0.3        & 6.3      & 0.8      & 0.002   \\
\ MSB23& 110  & 240  & 6.7     & 2.8 & 2.55       & 6.2      & 0.8      & 0.025   \\
\ MSB24& 90   & 200  & 6.5     & 1.6 & 2.55       & 6.2      & 0.8      & 0.010   \\
\ MSB25& 100  & 220  & 5.2     & 1.1 & 0.8        & 6.6      & 1.2      & 0.005   \\
\ MSB26& 120  & 220  & 6.5     & 1.6 & 1.45       & 6.3      & 0.8      & 0.013   \\
\ MSB27& 80   & 230  & 6.4     & 0.7 & 0.35       & 6.5      & 0.8      & 0.002   \\
\ MSB28& 90   & 200  & 6.6     & 1.5 & 2.2        & 6.3      & 0.8      & 0.008   \\ \hline

\end{tabular}}

1: in \kms; 2: in \cm3 ; 3: in 10$^4$K; 4: in 0.01 units ;5: in 10$^{16}$cm; 6: in 10$^{-4}$ units; 7: in \erg ;

\end{table}

\begin{table}
\centering
\caption{Description of models Mpl1-Mpl28}
\small{
\begin{tabular}{lccccccccccc} \hline  \hline

   mod   & \Vs  &  \n0 & $F$   & $D$        & O/H      &   Ne/H   & \Hb \\
\        & 1    & 2    & 3     & 4          & 5         & 5        & 6        \\ \hline
\ Mpl1   & 160  & 260  & 3     &1300        & 6.6      & 0.7      & 0.12     \\
\ Mpl2   & 100  & 600  & 7.4   & 300        & 6.2      & 0.6      & 0.22      \\
\ Mpl3   & 100  & 600  & 60    & 1500       & 6.8      & 0.6      & 0.89     \\
\ Mpl4   & 100  & 1600 & 250   & 3          & 2.6      & 0.1      & 1.58     \\
\ Mpl5   & 100  & 1000 & 10    & 2          & 3.9      & 0.3      & 0.18     \\
\ Mpl6   & 100  & 400  & 4     & 590        & 6.5      & 0.5      & 0.15     \\
\ Mpl7   & 100  & 250  & 18    & 4000       & 6.7      & 0.4      & 0.27     \\
\ Mpl8   & 100  & 1100 & 70    & 400        & 6.2      & 0.2      & 1.26    \\
\ Mpl9   & 100  & 270  & 18    & 400        & 6.7      & 0.4      & 0.31   \\
\ Mpl10  & 100  & 400  & 9     & 950        & 6.5      & 0.5      & 0.23    \\
\ Mpl11  & 80   & 320  & 3.6   & 15         & 2.2      & 0.2      & 0.05     \\
\ Mpl12  & 100  & 380  & 6     & 17         & 2.2      & 0.2      & 0.08     \\
\ Mpl13  & 100  & 600  & 2.9   & 3.3        & 4.3      & 0.3      & 0.63     \\
\ Mpl14  & 200  & 370  & 6     & 320        & 5.0      & 0.4      & 0.15    \\
\ Mpl15  & 90   & 1400 & 74    & 400        & 6.5      & 0.5      & 1.5    \\
\ Mpl16  & 100  & 270  & 11    & 4500       & 6.7      & 0.3      & 0.4    \\
\ Mpl17  & 140  & 550  & 80    & 900        & 6.5      & 0.5      & 1.0   \\
\ Mpl18  & 100  & 600  & 7     & 340        & 6.2      & 0.6      & 0.23  \\
\ Mpl19  & 140  & 790  & 95    & 600        & 6.4      & 0.2      & 1.57   \\
\ Mpl20  & 100  & 850  & 99    & 940        & 6.5      & 0.3      & 1.41   \\
\ Mpl21  & 110  & 840  & 8.7   & 120        & 6.6      & 0.6      & 0.28   \\
\ Mpl22  & 110  & 800  & 8.7   & 120        & 6.9      & 0.4      & 0.26   \\
\ Mpl23  & 100  & 1000 & 14    & 110        & 6.6      & 0.4      & 0.37   \\
\ Mpl24  & 100  & 1000 & 12    & 110        & 6.5      & 0.4      & 0.35   \\
\ Mpl25  & 100  & 260  & 1.8   & 1300       & 6.6      & 0.7      & 0.12   \\
\ Mpl26  & 100  & 260  & 4.5   & 1000       & 4.6      & 0.25     & 0.1    \\
\ Mpl27  & 80   & 330  & 3     & 15.5       & 2.2      & 0.25     & 0.2    \\
\ Mpl28  & 100  & 1000 & 12    & 110        & 6.9      & 0.5      & 0.35   \\ \hline

\end{tabular}}

1: in \kms; 2:  in \cm3; 3: in 10$^{10}$ ph cm$^{-2}$ s$^{-1}$ eV$^{-1}$ at the Lyman limit; 4: in10$^{16}$cm; 5: in
10$^{-4}$ units; 6: in\erg ;

\end{table}

\begin{table}
\centering
\caption{Description of models Mpl01-Mpl028}
\small{
\begin{tabular}{lccccccccccc} \hline  \hline

   mod    & \Vs  &  \n0 & $F$   & $D$        & O/H      &   Ne/H   & \Hb \\
\         &  1   &  2   & 3     & 4          & 5        & 5       &      6    \\ \hline
\  Mpl01  & 160  & 260  & 3.3   & 11.3       & 6.6      & 0.7      & 0.156    \\
\  Mpl02  & 110  & 800  & 9.0   & 2.8        & 6.5      & 0.5      & 0.35    \\
\  Mpl03  & 100  & 230  & 0.9   & 20.0       & 6.8      & 0.8      & 0.033   \\
\  Mpl04  & 100  & 8000 & 1.    & 0.005      & 4.       & 0.1      & 0.018   \\
\  Mpl05  & 100  & 2300 & 4.    & 0.048      & 6.2      & 0.3      & 0.015   \\
\  Mpl06  & 100  & 100  & 2.    & 290        & 4.6      & 0.6      & 0.068   \\
\  Mpl07  & 100  & 580  & 0.6   & 0.35       & 3.3      & 0.3      & 0.003   \\
\  Mpl08  & 100  & 2800 & 5.8   & 0.038      & 4.9      & 0.1      & 0.02    \\
\  Mpl09  & 100  & 210  & 4.0   & 73.        & 3.7      & 0.6      & 0.15    \\
\  Mpl010 & 120  & 200  & 4.1   & 48.        & 5.0      & 0.5      & 0.13    \\
\  Mpl011 & 100  & 280  & 4.4   & 36.        & 5.7      & 0.7      & 0.13    \\
\  Mpl012 & 100  & 360  &  6    & 31.7       & 6.0      & 0.7      & 0.02    \\
\  Mpl013 & 90   & 270  & 6     & 90.        & 6.3      & 0.7      &  0.25   \\
\  Mpl014 & 190  & 200  & 9     & 40         & 6.6      & 0.7      & 0.3     \\
\  Mpl015 & 170  & 700  & 15    & 10.        & 6.4      & 0.7      & 0.51    \\
\  Mpl016 & 150  & 210  & 8     & 76.        & 5.5      & 0.5      & 0.47    \\
\  Mpl017 & 110  & 250  & 11    & 110.       & 5.8      & 1.0      & 0.45    \\
\  Mpl018 & 110  & 500  & 3.9   & 4.5        & 6.6      & 0.8      & 0.14    \\
\  Mpl019 & 130  & 640  & 27    & 16.        & 6.0      & 0.3      & 1.09    \\
\  Mpl020 & 150  & 250  & 3.5   & 14.2       & 5.0      & 0.7      & 0.16    \\
\  Mpl021 & 100  & 580  & 7     & 7.5        & 6.2      & 0.8      & 0.29    \\
\  Mpl022 & 110  & 900  & 12    & 2.25       & 3.8      & 0.3      & 0.29    \\
\  Mpl023 & 140  & 840  & 13    & 1.9        & 6.6      & 0.5      & 0.40    \\
\  Mpl024 & 110  & 900  & 12    & 2.7        & 5.4      & 0.5      & 0.4     \\
\  Mpl025 & 70   & 230  & 2.1   & 450        & 6.9      & 0.9      & 0.1      \\
\  Mpl026 & 90   & 240  & 2.1   & 47         & 7.6      & 0.8      & 0.096   \\
\  Mpl027 & 90   & 240  & 2.5   & 47         & 6.6      & 0.8      & 0.088   \\
\  Mpl028 & 120  & 900  & 10    & 1.6        & 3.4      & 0.3      & 0.19    \\  \hline

\end{tabular}}

  1: in \kms; 2: in \cm3; 3: in 10$^{10}$ ph cm$^{-2}$ s$^{-1}$ eV$^{-1}$ at the Lyman limit; 4: in 10$^{16}$cm; 
5: in 10$^{-4}$; 6: in \erg

\end{table}

In Table 1 we compare the reddening corrected observed line ratios  to \Hb with model results.
Each observed spectrum  is recognizable by its ID number (Ly et al 2015). 
 We refer to the corrected observed [OIII]5007+/\Hb line ratios which appear in parenthesis
(last column),  calculated by 
([OIII]5007+/[OIII]4363)$_{corr}$$\times$([OIII]4363/\Hb)$_{corr}$, because  the [OIII]5007+/\Hb line ratios
 reported in Table 1
 from  Ly et al.  (2015, table 2) were not corrected.  The observed corrected \RO3 ([OIII]5007+/[OIII]4363)
 line ratios  (C.Ly 2015, private communication)
appear in Table 1 last column.

The models  are constrained by the FWHM of the line profiles 
(C.Ly, 2015, private communication) 
which  roughly indicate the   velocity
field of the emitting gas and give an initial hint  of the shock velocity. 

Models MSB1-MSB28 refer to  photoionization  by the SB (+shocks)  of clouds 
   propagating outwards from   the galaxy.
Models Mpl1-Mpl28 account for photoionization
by an AGN + shocks. The clouds propagate outwards from  the AGN. Models  Mpl01-Mpl028  refer 
to photoionization by an AGN  + shocks,  but the clouds
propagate towards the active nucleus.
Models MSB1-MSB28, Mpl1-Mpl28 and Mpl01-Mpl028 are described in Tables 2, 3 and 4, respectively.
The  physical conditions and the relative abundances  presented in  Tables 2-4 are regarded
as the results of modelling for each galaxy.
We have selected the models showing  discrepancies for the strongest line ratios within 10\% and for the weakest ones
(e.g. [OIII]4363/\Hb) within 50\%.  However, in a few cases (e.g. for models MSB23 and MSB24) we could 
not find a good  approximation of calculated  to observed  [OIII]4363/\Hb without destroying the good fit
of [OIII]5007+/\Hb and [OII]3727+/\Hb.

The set of parameters adopted to model the SB  galaxies (Table 2) shows pre-shock densities, SB effective temperatures
and ionization parameters in agreement with those calculated in this redshift range for different types of galaxies
(Contini 2015).
The geometrical thickness of the emitting clouds are at the lower limit. The O/H and Ne/H relative
abundances are close to solar.
The parameters of AGN dominated models in the outflow case (Table 3) which yield the satisfactory fit of the
observed line ratios are  dominated by relatively high  \n0 and   $D$  suitable to the
narrow line emission region.
Moreover,  O/H are close to solar for most of the objects, 
but 0.3 solar for Mpl4, Mpl11, Mpl12 and Mpl27. Ne/H are lower than solar
(10$^{-4}$) by a factor $>$ 2 for most galaxies. However, 
 the models refer to the doublet and the data to the deblended line.
  Ne  corresponds to the  1s2, 2s2, 2p6 closed atomic configuration, therefore Ne is less adapted to 
link with other species inside dust grains. 

The parameters selected to reproduce the observed spectra for AGN models in the inflow case (Table 4),
 show relatively high densities, a few large clouds and O/H and Ne/H slightly lower than
solar, except for Mpl04, Mpl07, Mpl09, Mpl022 and Mpl028 where O/H are $\sim$ half  solar.
High preshock densities in the  inflow case are reasonable, considering that the IS clouds
in the surrounding of the AGN are accreted toward the AGN.

In Fig. 3  we compare  the modelling results  with the Kewley et al (2001) diagnostics.
  [OI], [NII] and [SII] lines were not observed, so we use the results of models which better fit
the observed line ratios. 
We adopt O/H solar, N/H =0.3 solar and S/H solar  (Contini 2015,  table 1).
Fig. 3 top diagram shows that SB models are definitively in the SB zone,
AGN models in the AGN zone, even if some models calculated for AGN galaxies
in the outflow case, are on the border line. For the other two diagrams ([OIII]/\Hb vs [NII]/\Ha and
[OIII]/\Hb vs [SII]/\Ha)  most of  the SB models are on the border line.
This already suggests that some galaxies show a complex character of SB and AGN.
The spread of the [OI]/\Ha line ratios and, in particular, the separation between the inflow from the
outflow results, is due to the geometrical thickness large range of the emitting clouds in the AGN
dominated models.
In the outflow case (Fig. 1), the clouds may contain in the internal region, between the shock front edge and the
edge illuminated by the AGN flux, a low temperature zone with large quantities of
gas in the physical conditions suitable  to the [OI] line.
The same could explain the spread of the [SII] lines.  However, when referring to the data instead of
to the models, the S/H relative abundance also changes
from object to object because S is easily trapped into dust grains.
The  [NII]/\Ha  line ratios  fill  a more compact region throughout the diagram 
and the AGN inflow and outflow models overlap,
because the region throughout the cloud where N$^+$  prevails  is radiation dominated.
We have found by detailed modelling  of different types of spectra emitted from  galaxies at  0.01$\leq$z$\leq$3
that N/H varies within a factor of 10 throughout all the z range (Contini 2016, fig. 5).
So inserting eventual observation data instead of model results in Fig. 2 middle diagram, the picture would
change.

In Fig. 4 we check the  modelling result precision. 
The observational uncertainties are not shown for sake of clarity.
The [OIII] 5007+/\Hb and [OII]3727+/\Hb calculated line ratios reproduce satisfactorily
the data, indicating that the physical conditions and the relative abundances,  adopted in the calculations
of the spectra, are sound. 
The same models were used to calculate for each object the   [OIII]4363/\Hb line ratios shown
in Fig. 4, right diagram.
The observed [OIII]4363 lines are  rather weak. Even accepting discrepancies  by a factor of 2,
 Fig. 4 (right diagram) indicates that several models referring to the AGN outflow 
case   should be dropped.

\section{Discussion and concluding remarks}

\subsection{Physical conditions,  metallicities and SFR  in the sample galaxies}

\begin{figure*}
\centering
\includegraphics[width=5.8cm]{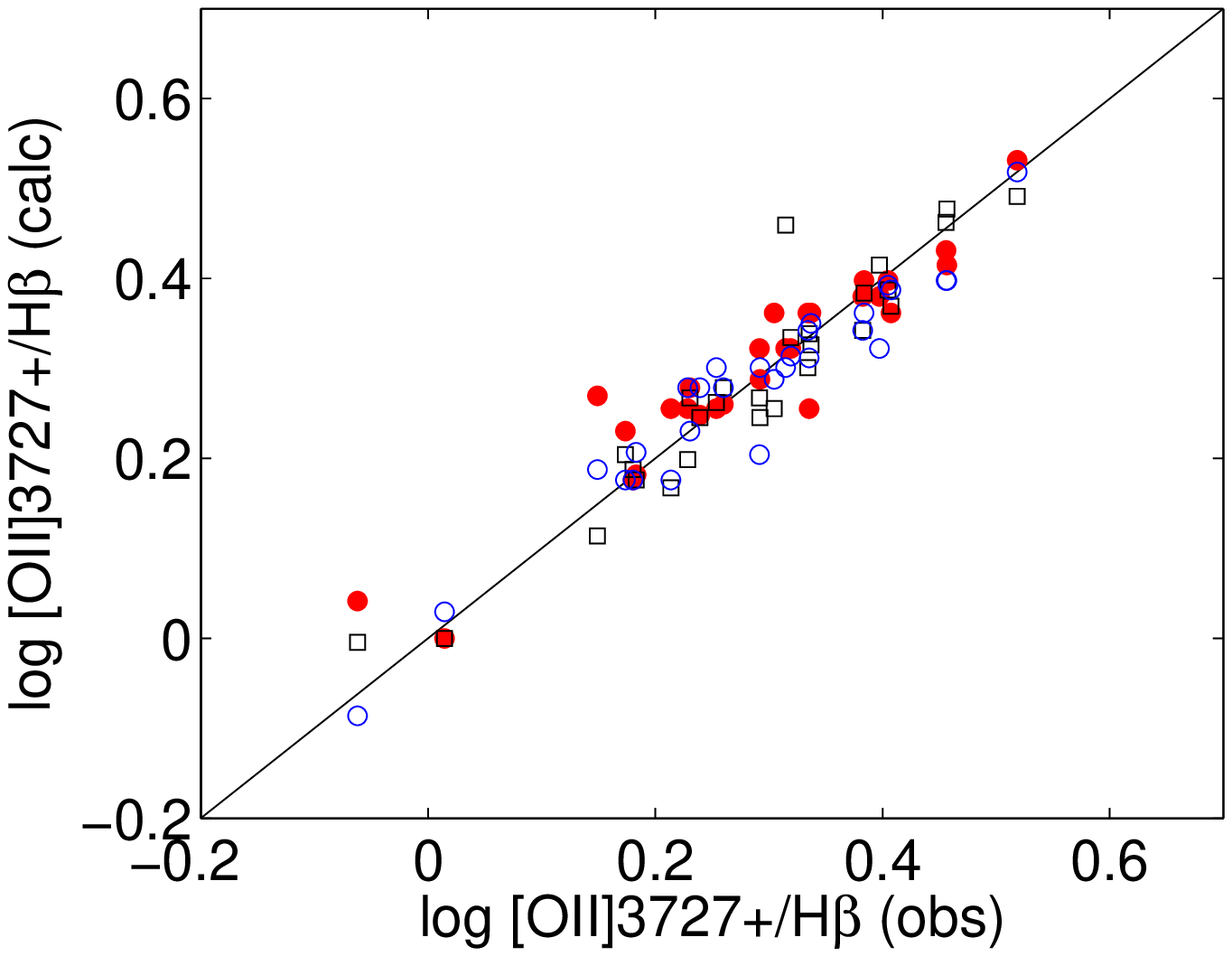}
\includegraphics[width=5.8cm]{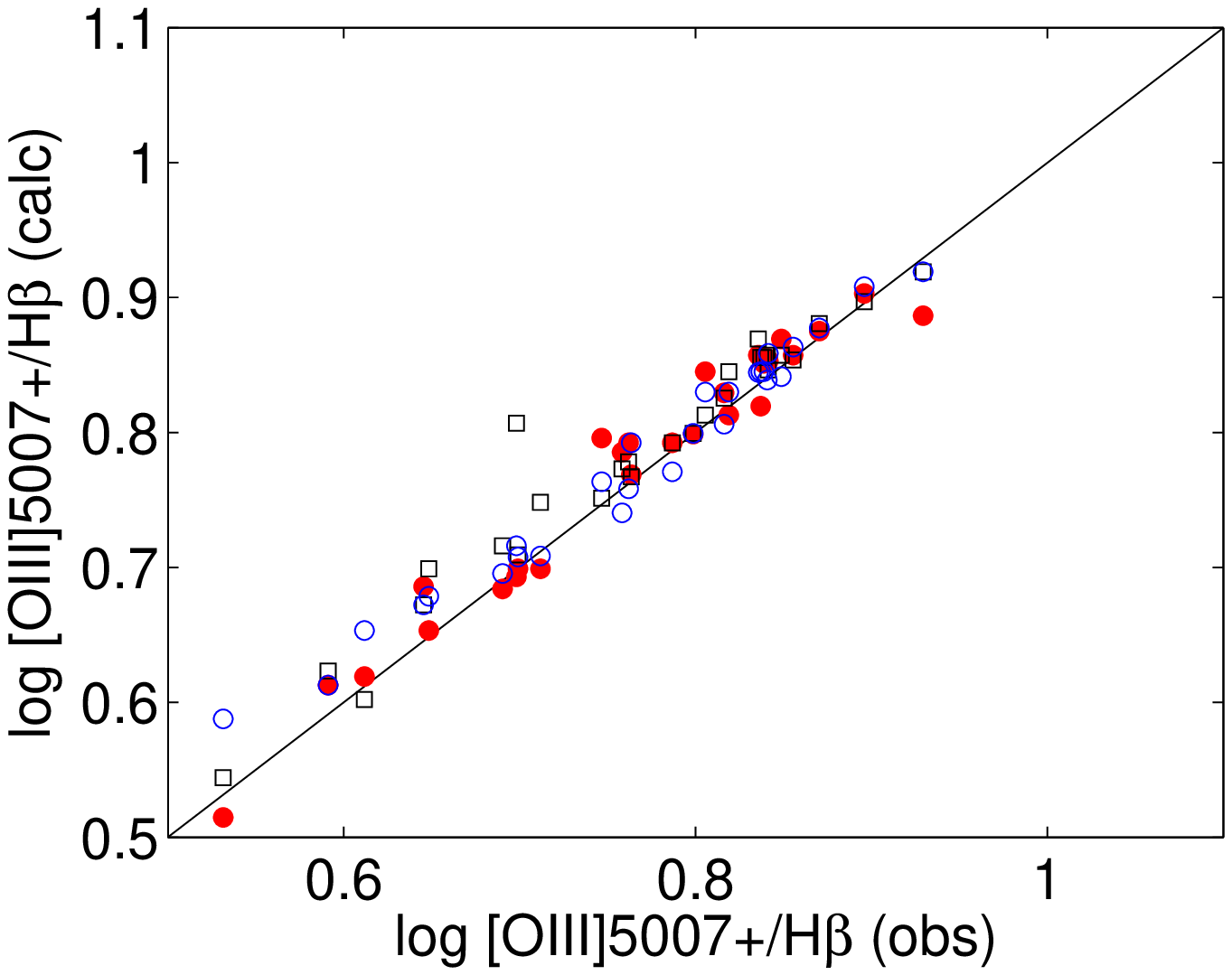}
\includegraphics[width=5.8cm]{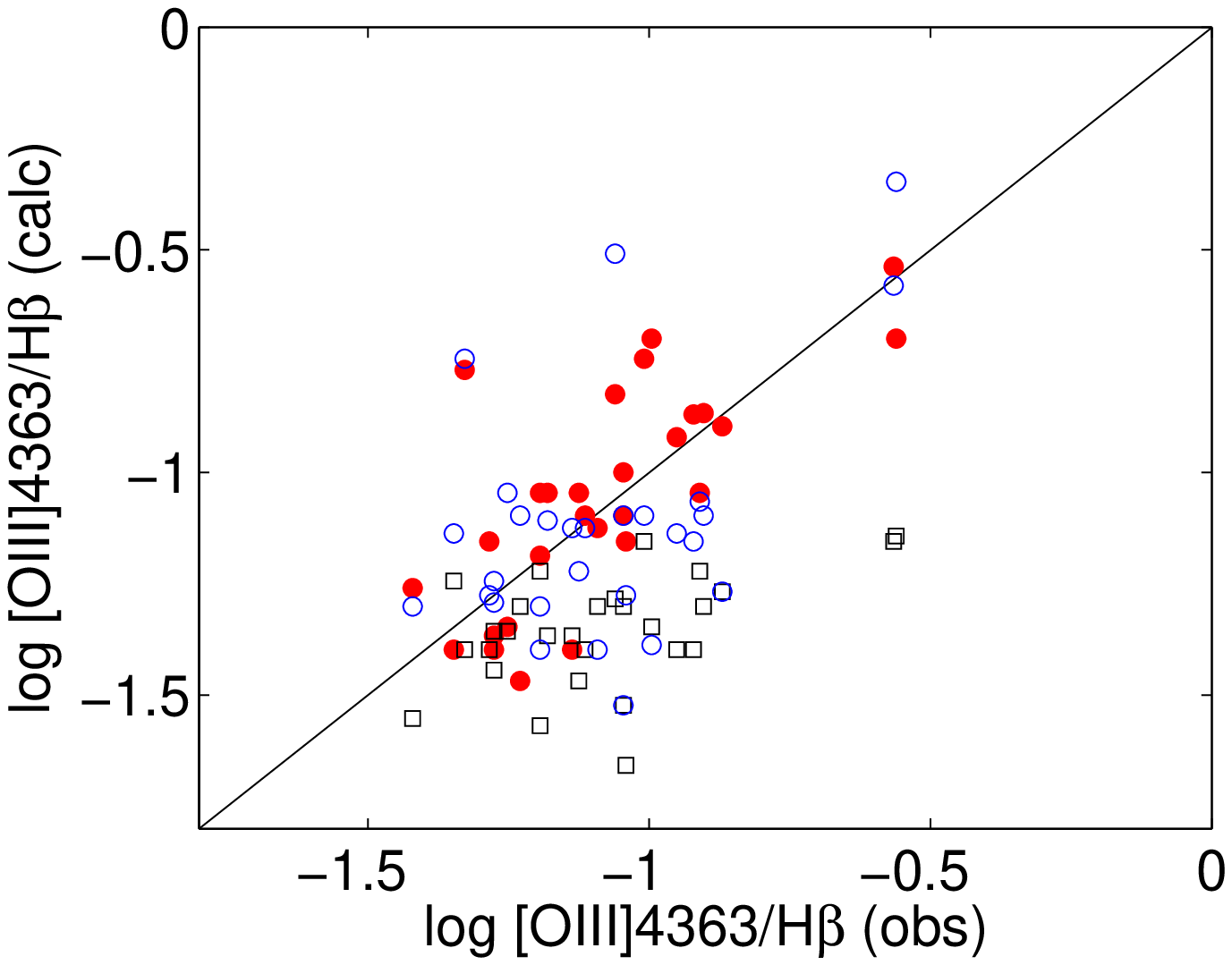}
\caption{
Left : observed vs calculated [OII]3727+/\Hb. Middle :observed vs calculated [OIII]5007+/\Hb.
Right : observed vs calculated [OIII]4363/\Hb for single galaxies. Symbols as in Fig. 3.
}
\end{figure*}

\begin{figure}
\centering
\includegraphics[width=9.4cm]{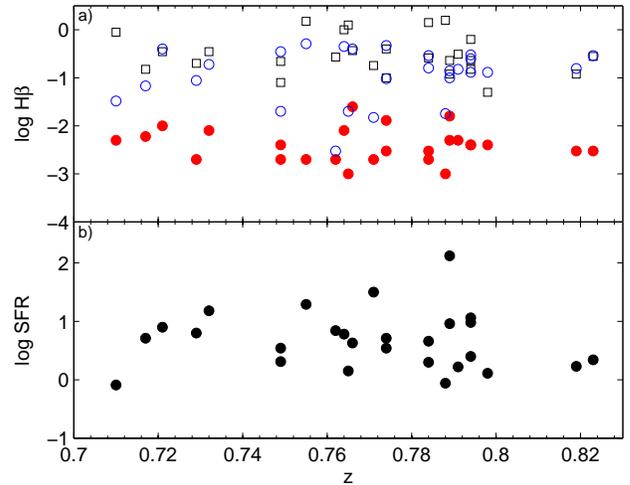}
\caption{
Top panel: calculated \Hb  in \erg. Symbols as in Fig. 4
Bottom panel : the SFR in \msol yr$^{-1}$    given by Ly et al.
}
\end{figure}

Tables 2-4  show that \Vs $\sim$ 100 \kms  and  \n0 $\sim$ 200 \cm3  in most of the galaxies
characterize the gaseous   clouds  
 in the SB environment, while   preshock densities are  higher 
by a factor of 1.5-10 in  the AGN clouds. 
The SB clouds  show major fragmentation
($D$ $\sim$ 10$^{16}$ cm). In the AGN,  
$D$ ranges from 10$^{16}$ cm   to  a maximum of  $>$10 pc for 
the ejected clouds. Geometrical thick clouds up to D=3 pc were calculated for the merger galaxy
NGC 6240 (Contini 2012b). 
Combining an average \Hb flux of 10$^{-16}$ \erg observed at Earth (Table 1)  with an average  \Hb flux calculated at the nebulae (Table 4)
for the SB clouds ($\sim$0.01 \erg) and for the clouds in the AGN environment ($\sim$0.1 \erg),
we obtain  the average distance of the emitting clouds from the SB R$_{SB}$$\sim$ 0.5 kpc and from the AGN
R$_{AGN}$$\sim$ 0.16 kpc, respectively,  adopting a filling factor $\sim$1.
 The starburst effective temperatures \Ts are similar to those found in
  quiescent galaxies (4-7 10$^4$ K, Contini 2014b). 

Metallicities, in term of the O/H relative abundances, are  a crucial issue
for  galaxies at  high z. 
 The  O/H relative abundances  calculated by detailed modelling are  about solar (6.6 10$^{-4}$, Allen 1976) 
for all the objects, 
except for a few  AGN with O/H= 0.3 -0.5 solar. 
Ly et al (2015) obtain  0.223-3.23 10$^{-4}$ by the T$_e$ based metallicity determination
(cf. Contini 2014a). The  discrepancies between metallicity results obtained by different modelling  methods are
 explained in Sect. 2.

 In Fig. 5 the \Hb  absolute fluxes (in \erg) calculated   at the emitting nebula and the
 SFR (in \msol yr$^{-1}$)  given  by Ly et al.  are  shown as function of the redshift.
Throughout a small z range, they do not show any trend, while (Contini 2016, fig. 6 and references therein)
SFR  in SN and GRB host galaxies were found to increase with z on a large scale for z$\geq$0.1.
At lower z, SFR do not show any particular slope. On the other hand, SFR presented by Ramos Almeida et al (2013) for 
X-ray and mid-infrared selected galaxies at 0.4$\leq$z$\leq$1.15 exhibit  an increasing trend.
More objects should be considered. 
 Moreover,   SFR $\propto$ L(\Ha)$_{Earth}$ (the \Ha luminosity  observed at Earth) and
L(\Ha)$_{Earth}$ = L(\Ha)$_{nebula}$ (the \Ha luminosity calculated  at the nebula) which is $\propto$ \Ha R$^2$. 
We adopt \Ha/\Hb$\sim$3. R is the distance of the emitting cloud from the radiation source.
So SFR $\propto$ \Hb, assuming  an average R. 
Ly et al results show that  SFR  in AGN and SB galaxies are very similar, whereas from the \Hb flux 
calculated at the nebulae (Fig. 5, top diagram), higher SFR for AGN  are  expected. 
This discrepancy  can be explained by the coexistence  of AGN and SB 
in most of the sample  objects.

\begin{figure*}
\centering
\includegraphics[width=8.8cm]{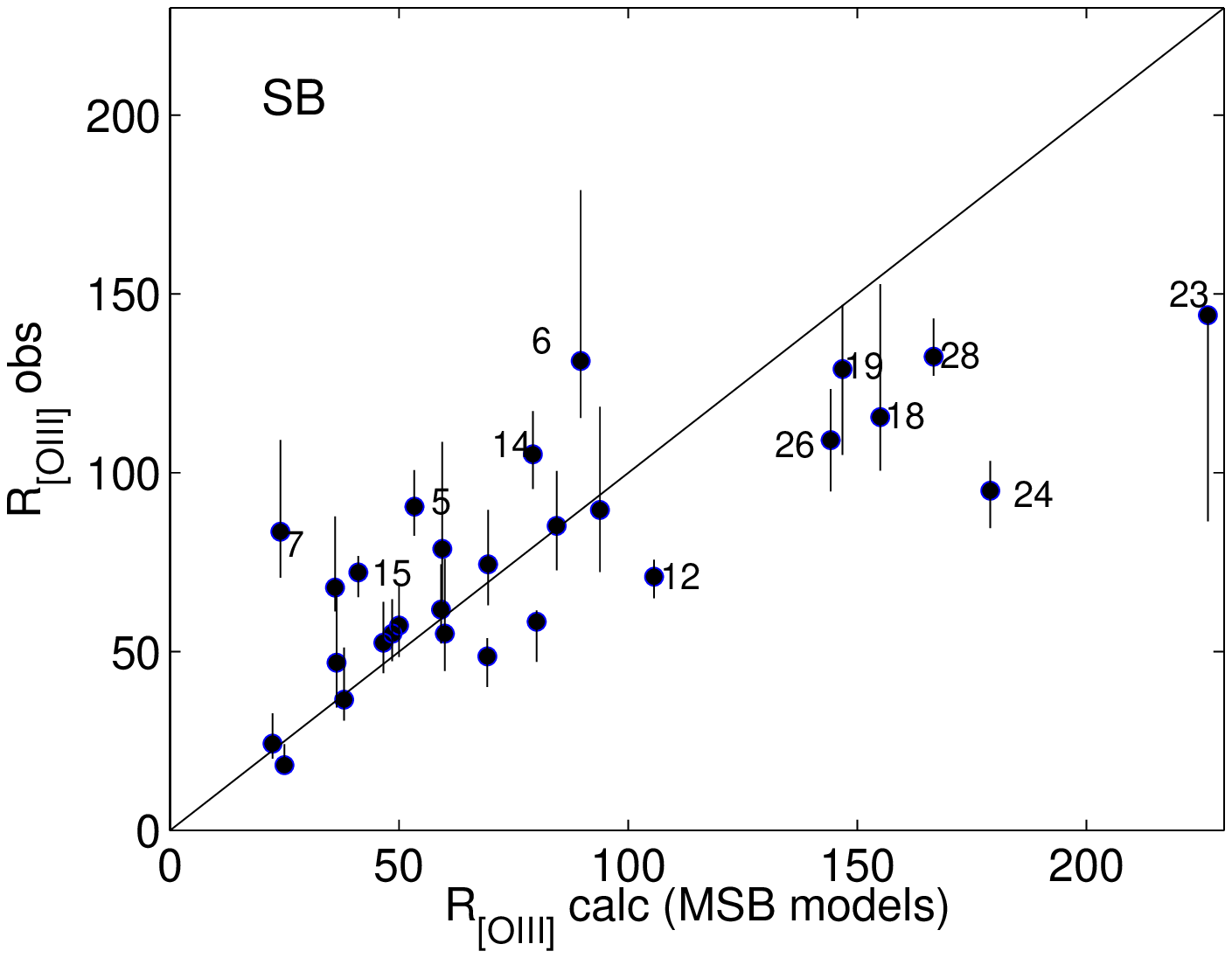}
\includegraphics[width=8.8cm]{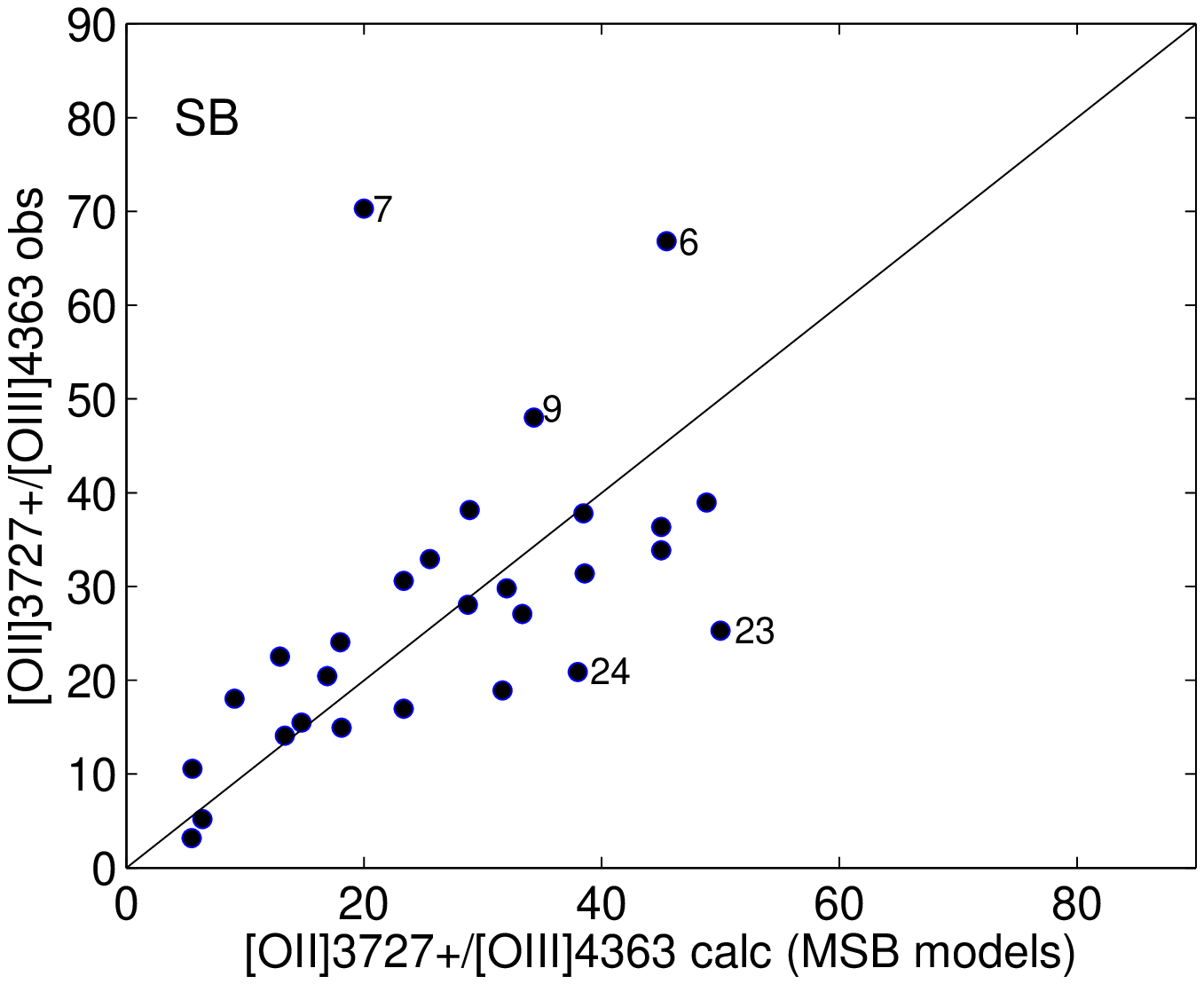}
\includegraphics[width=8.8cm]{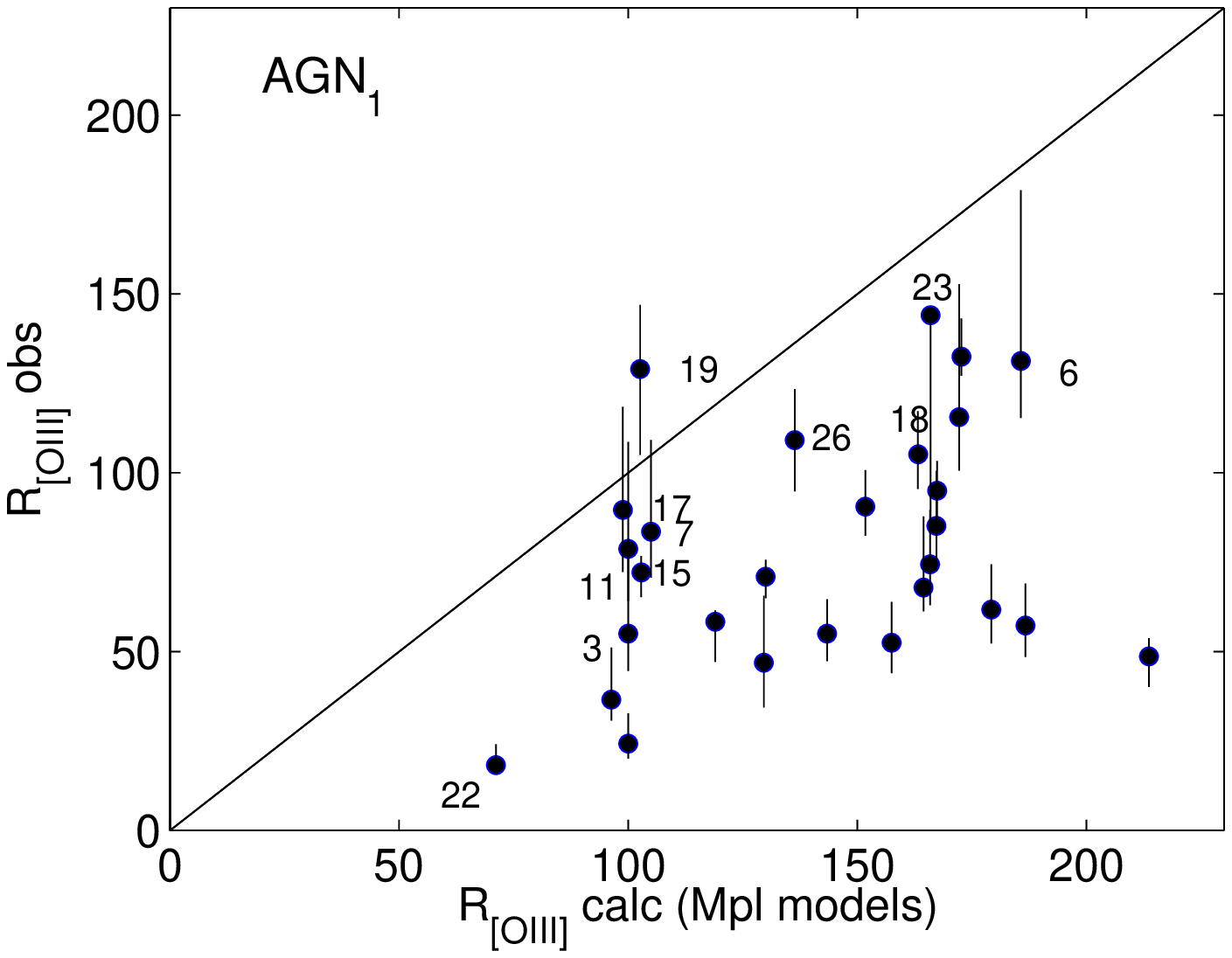}
\includegraphics[width=8.8cm]{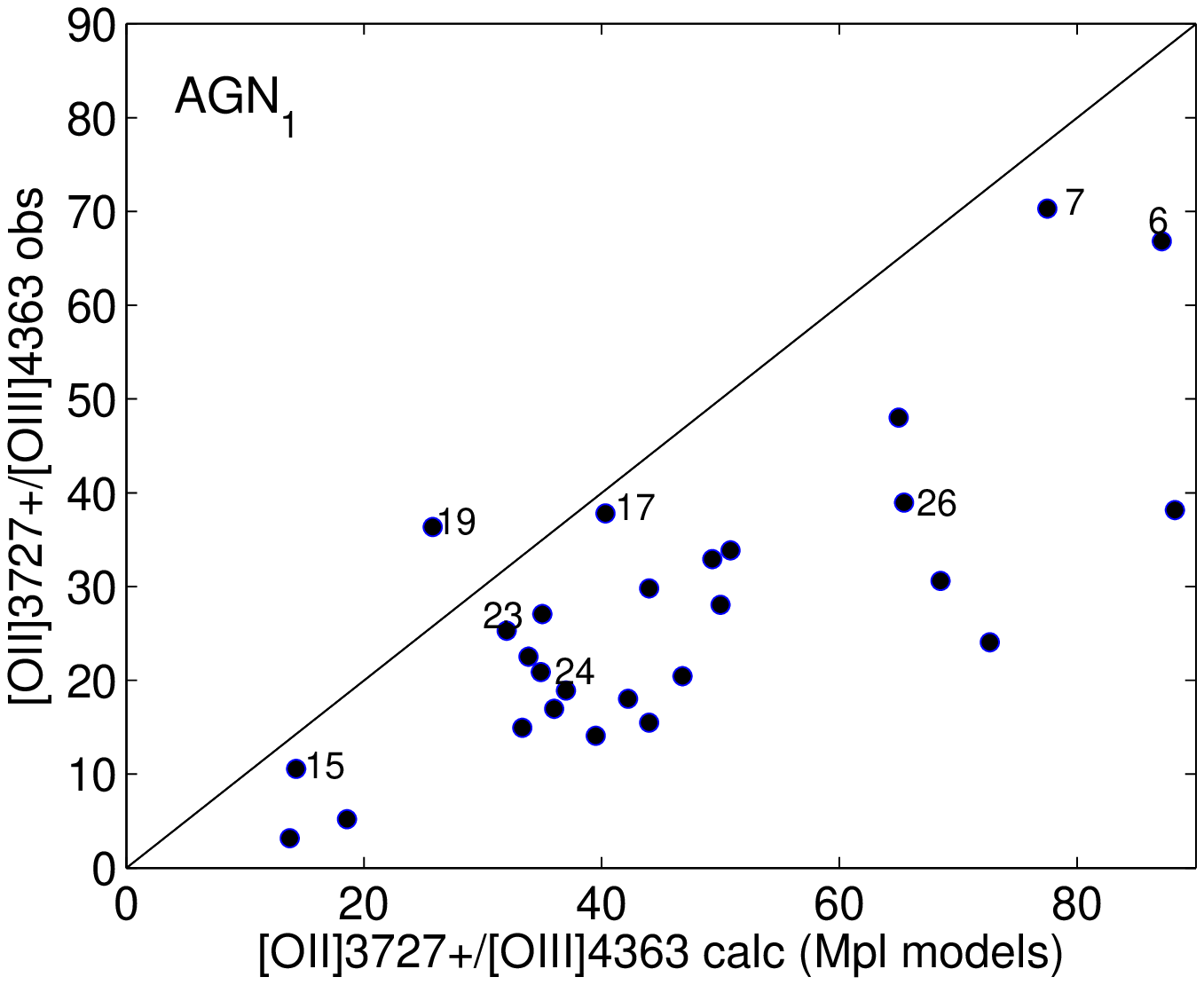}
\includegraphics[width=8.8cm]{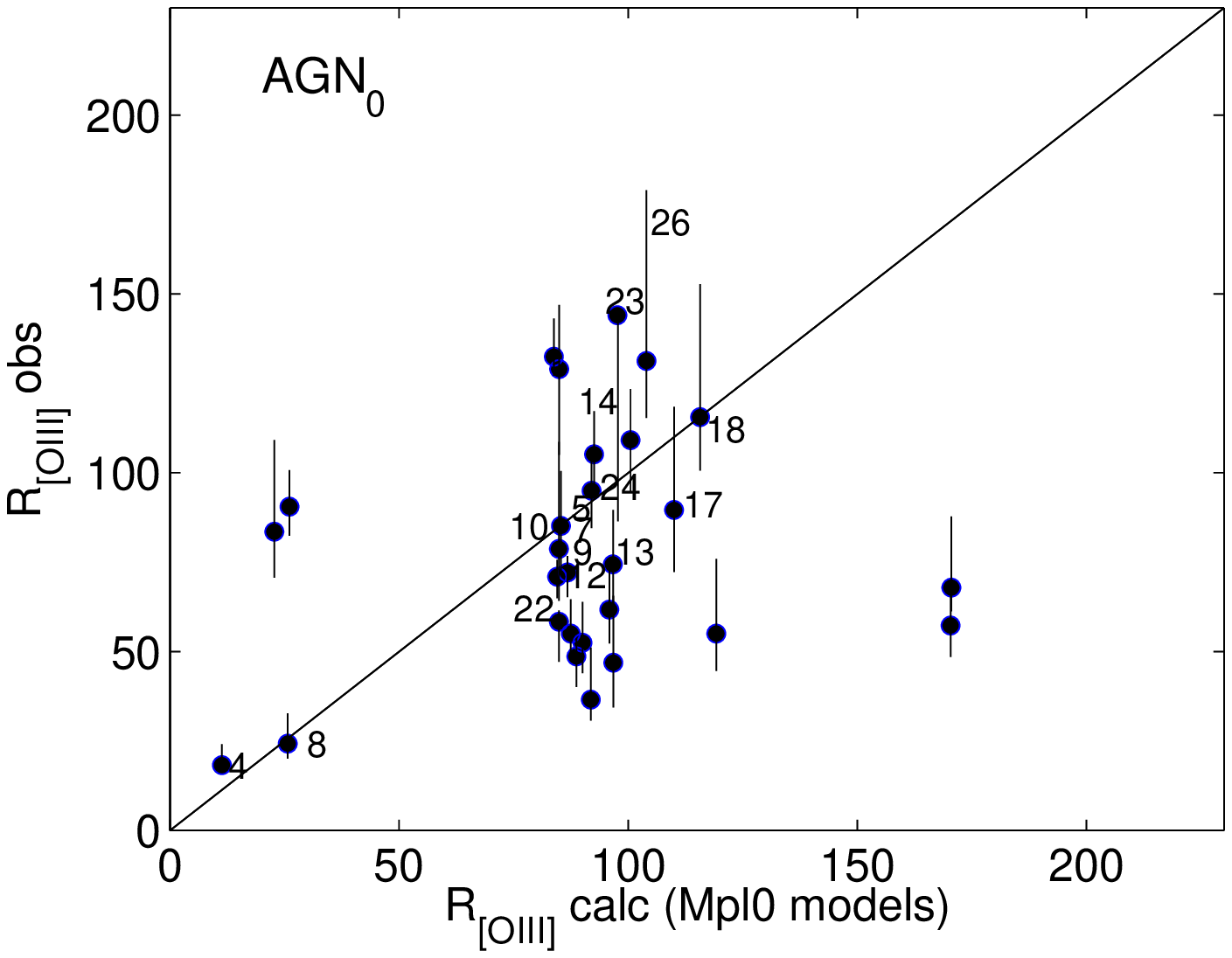}
\includegraphics[width=8.8cm]{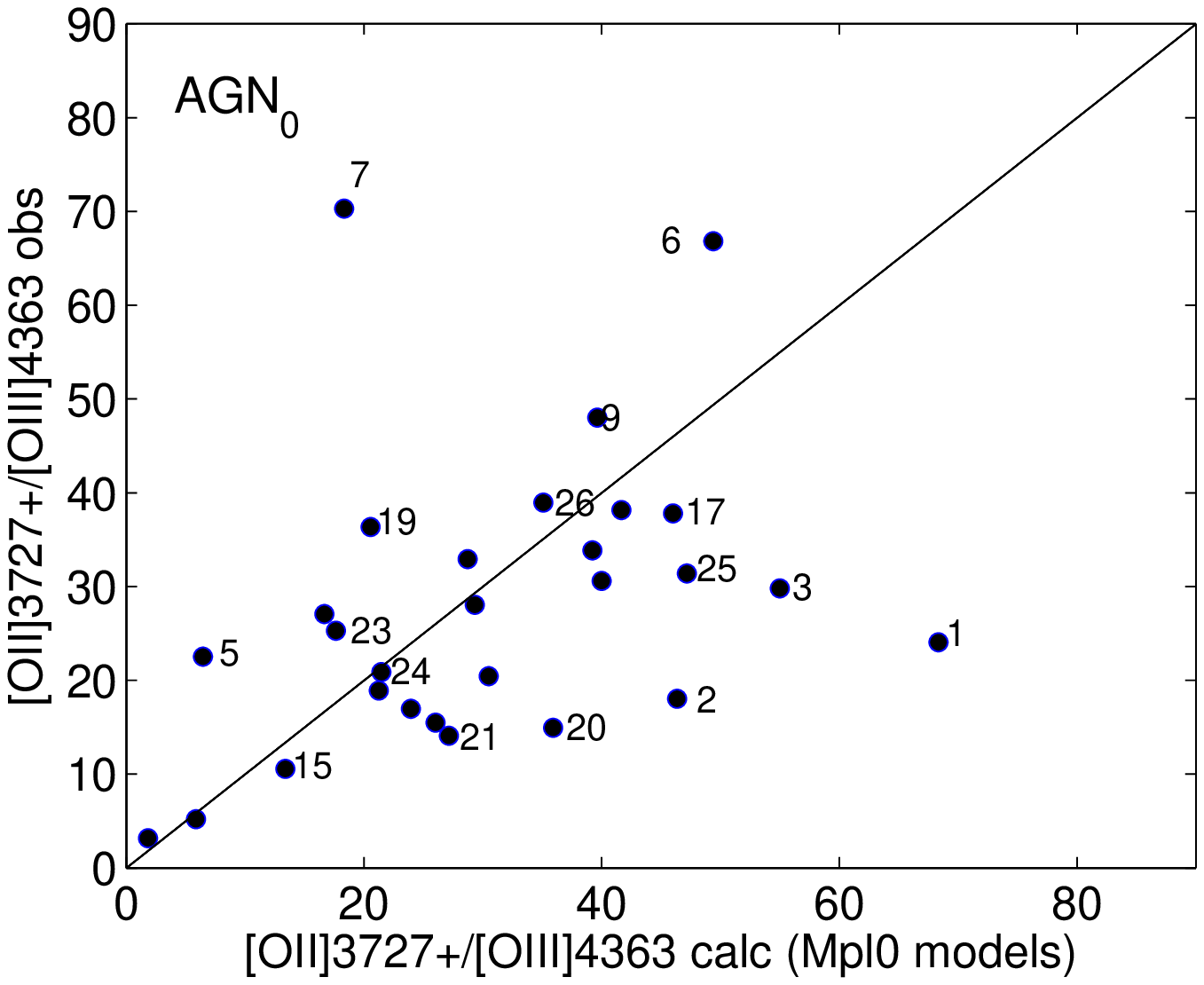}
\caption{
Observed vs calculated  \RO3 (left) and [OII]3727+/[OIII]4363 (right) for single galaxies.
ID numbers (see Table 1).
}
\end{figure*}

\begin{table*}
\centering
\caption{Galaxy types resulting from the modelling}
\tiny{
\begin{tabular}{lccccccccccccccccccccccccccccc} \hline  \hline
 1& 2& 3& 4& 5& 6& 7& 8& 9& 10& 11& 12& 13& 14&15&16&17&18&19&20&21&22&23&24&25&26&27&28 \\ \hline
 S&S &S &S&S & - & -& S&-& S& S& S& S& S&S&S&S&S&S&S&S&S&-&-&S&S&S&S \\
 - & -&1 &- &- &1&1&- &- &- & - & - & - & - &1& -&1&- &1&- &- &- &1&- &- &-&- &-  \\
 - & -&- &0&- &- &- &0&0&0& - & 0& 0 & 0&- & -&0&0& -&- &- &0 &0&0&- &0& - &-  \\ \hline
\end{tabular}}
\end{table*}

\subsection{Multiple radiation sources in  single galaxies}

By AGN and SB dominated models,  
we reproduced successfully  the [OIII]5007+/\Hb and [OII]/\Hb observed line ratios, not always the
[OIII]4363/\Hb  from each galaxy.
We  investigate   whether the photoionization
source  in  each  object is an AGN, a SB or both,   comparing in
Fig. 6   left diagrams and right diagrams,  the calculated with observed  corrected \RO3 and [OII]/[OIII]4363, respectively.
The top  diagrams refer to  results obtained  by SB dominated models,
the middle ones  by AGN dominated models  with outflowing  clouds and the bottom diagrams show the results of
 models  dominated by  AGNs which accrete the surrounding clouds.
Fig. 6 shows that
models MSB1-MSB28 (Table 2)
reproduce the data within the observational  errors,
except  for ID 6, 7, 23 and 24 (Fig. 6 top  diagram).
Most of  the  Mpl1-Mpl28 model results (Table 3)
overpredict the data, except ID 19 (Fig. 6 , middle).
 ID 3, 6, 7, 15, 17, 19 and 23 \RO3  are reproduced  within $\sim$ 30\%.
Fig. 6 bottom  diagram refers to models Mpl01-Mpl028 (Table 4).
About half of the object spectra are well reproduced by the AGN accretion models.
The results show that some   line ratios can be  explained with the same precision by both the
SB and AGN dominated models.
  The galaxy types are schematically shown in Table 5 (S=SB, 1=AGN$_1$, 0=AGN$_0$),
 where AGN$_1$ refer to ejection of matter outwards and  AGN$_0$ to accretion.

 Summarizing,  calculation results suggest that  most of the objects  contain a SB, 
about 5 objects are AGNs.
Half of the  galaxies show multiple radiation sources : a SB + an  accreting AGN. 
ID 3, 15, 17 and  19
show an SB + an AGN with outflowing  matter.  ID 17 and ID 23   host a double AGN, but ID
17   also a SB. 
The AGNs   show gas accretion  rather than  outflow, suggesting
 an AGN-SB correspondence.
Multiple nuclei are found in local galaxies which  derive from  merging, e.g. in Arp 200 at z=0.018 (Graham et al 1990),
where O/H relative abundances are about solar and N/H are higher than solar by a factor of $\sim$1.5 
throughout the starburst region (Contini 2013).
To complete our investigation we need  measurements of the [OI], H$\alpha$, [NII]  and [SII] emission lines,
which are only available with near-infrared spectroscopy at z$\sim$0.8.

\section*{Acknowledgements}
I am  grateful to the referee for valid comments which improved the presentation of the paper.
I  thank Dr. Chun Ly for preparing the spectra in the suitable format
and for helpful discussions.

\section*{References}

\def\ref{\par\noindent\hangindent 18pt}
\ref Allen, C.W. 1976 Astrophysical Quantities, London: Athlone (3rd edition)
\ref Anders, E., Grevesse, N. 1989, Geochimica et Cosmochimica Acta, 53, 197
\ref Asplund, M., Grevesse, N., Sauval, A.J., Scott, P. 2009, ARAA, 47, 481
\ref Blank, M. , Duschl, W.J. 2014, IAUS, 303, 379
\ref Contini, M. 2016, MNRAS, in press (ArXiv:1605,02438)
\ref Contini, M. 2015, MNRAS, 452,3795
\ref Contini, M. 2014b, A\&A, 572, 65
\ref Contini, M. 2014a, A\&A, 564, 19
\ref Contini, M. 2013, MNRAS, 429, 242
\ref Contini, M., Contini, T. 2007, AN, 328, 953
\ref Contini, M., Aldrovandi, S.M.V. 1986, A\&A, 168, 41
\ref Dixon, T.G., Joseph, R.D.  2011, ApJ, 740, 99 
\ref Engel, H. et al. 2010, A\&A, 524 ,56
\ref Graham G. R., Carico D. P., Matthews K., Neugebauer G., Soifer B. T., Wilson T. D. 1990,
ApJ, 354, L5.
\ref Hopkins, P.F. , Quataert, E., Murray, N.  2011, MNRAS, 417, 950
\ref Kewley, L.J., Dopita, M.A., Sutherland, R.S., Heisler, C.A., Trevena, J. 2001, ApJ, 556, 121
\ref Ly, C., Rigby, J.R., Cooper, M., Yan, R.  2015, ApJ, 805, 45
\ref Osterbrock, D.E. 1974 in 'Astrophysics of Gaseous Nebulae' W.H. Freeman and Company, San Francisco
\ref Ramos Almeida, C., Rodr\'{i}guez Espinosa, J. M., Acosta-Pulido, J. A., Alonso-Herrero, A., Pérez García, A. M., Rodr\'{i}guez-Eugenio, N.
2013, MNRAS, 429, 3449
\ref Zubovas, K., Nayakshin, S., King, A., Wilkinson, M. et al 2013, MNRAS, 433, 3079

\end{document}